\documentclass[twocolumn]{aastex631}
\usepackage{amsmath,amsfonts,amssymb,graphicx,chngcntr,multirow, float,booktabs, afterpage}
\usepackage[]{hyperref}
\usepackage{etoolbox}
\patchcmd{\endthebibliography}{\newpage}{}{}{\typeout{WARNING: thebibliography \newpage patch failed}}
\hypersetup{colorlinks=true}

\shorttitle{QPE spectral-timing}

\begin{document}

\title{Quasi-periodic eruption spectral-timing: EMRIs crossing warped accretion disks}

\author[0000-0002-0568-6000]{Joheen Chakraborty}
\thanks{Both authors contributed equally to this work.}
\affiliation{Department of Physics \& Kavli Institute for Astrophysics and Space Research, Massachusetts Institute of Technology, Cambridge, MA, 02139, USA}
\email{joheen@mit.edu}

\author[0009-0006-4590-8583]{Andrew Mummery} 
\thanks{Both authors contributed equally to this work.}
\affiliation{School of Natural Sciences, Institute for Advanced Study, 1 Einstein Drive, Princeton, NJ 08540, USA}
\email{amummery@ias.edu}

\author[0000-0001-9185-5044]{Eliot Quataert}
\affiliation{Department of Astrophysical Sciences, Princeton University, Peyton Hall, Princeton, NJ 08540, USA}

\author[0000-0003-4054-7978]{Riccardo Arcodia}
\affiliation{Black Hole Initiative at Harvard University, 20 Garden Street, Cambridge, MA 02138, USA}

\author[0000-0002-8304-1988]{Itai Linial}\thanks{NASA Einstein Fellow}
\affiliation{Center for Cosmology and Particle Physics, Physics Department, New York University, New York, NY 10003, USA}

\author[0000-0003-0172-0854]{Erin Kara}
\affiliation{Department of Physics \& Kavli Institute for Astrophysics and Space Research, Massachusetts Institute of Technology, Cambridge, MA, 02139, USA}

\begin{abstract}
    Quasi-periodic eruptions (QPEs) are repeating X-ray transients from galactic nuclei potentially caused by a stellar-mass orbiter repeatedly colliding with the accretion disk of a supermassive black hole (SMBH) in an extreme mass-ratio inspiral (EMRI). In this picture, the eruption arrival times encode the orbit and its relativistic precessions, while the quiescent spectral energy distribution (SED) probes the same disk and SMBH, providing two independent constraints on the same orbit-disk system. We present a joint spectral-timing framework that takes advantage of this by simultaneously fitting an analytic QPE timing model built on Kerr geodesic frequencies and a warped disk geometry, coupled to a self-irradiated disk SED model sharing the same SMBH mass, spin, and disk parameters. We apply it to the QPE sources GSN~069 and eRO-QPE2, showing that the QPE timings and quiescent spectra can be reproduced simultaneously in each, providing circumstantial support for the EMRI model and warped disks in these systems. However, given the sparse timing data in both sources, neither admits a unique solution. In eRO-QPE2, we identify six physical modes which reproduce the timings and SED comparably well, motivating denser observing campaigns to identify one unambiguously. Our results demonstrate both the promise of QPE spectral-timing as a precision probe of SMBHs, their accretion disks, and orbiting companions, and the aliasing subtleties of sparsely sampled QPE timings.
\end{abstract}

\section{Introduction} \label{sec:intro}

Quasi-periodic eruptions (QPEs) are soft thermal X-ray flares that recur every few hours to days in the nuclei of low-mass galaxies, with thirteen systems currently known \citep{Miniutti19,Giustini20,Arcodia21,Arcodia24a,Arcodia25,Chakraborty21,Chakraborty25a,Quintin23,Nicholl24,Hernandez25a,Baldini26}. Several sources have shown QPEs emerging from the accretion disks formed in recent tidal disruption events (TDEs) \citep[e.g.][]{Chakraborty21,Chakraborty25a, Miniutti23a, Quintin23, Nicholl24, Baldini26}, and some theoretical models attribute these eruptions to a stellar-mass secondary (a star or compact object) on an inclined, mildly eccentric orbit repeatedly striking the disk \citep[e.g.][]{Xian21,Franchini23,Linial23b,Linial25,Tagawa23,Mummery25a,Yao25a}. In this picture, QPEs are the first observed counterparts to extreme mass-ratio inspirals (EMRIs) at $\mathcal{O}(100\,r_g)$ around SMBHs, where $r_g\equiv GM_\bullet/c^2$ is the gravitational radius of an SMBH of mass $M_\bullet$.

If the orbiter-disk collision picture is correct, the eruption arrival times are a rich observable encoding the physical properties of the SMBH, orbiter, and disk via relativistic apsidal (Schwarzschild) and nodal (Lense--Thirring) precession, which modulate the crossing times over super-periods of tens to thousands of orbits. Timing campaigns have indeed detected super-orbital modulations in multiple sources \citep{Chakraborty24,Chakraborty26,Miniutti25,Arcodia26}, and some studies have begun to explore orbit-fitting of individual QPE systems \citep{Zhou24a,Zhou24b,Chakraborty25c}. However, existing timing data are sparse: brief epochs separated by long gaps miss many cycles, making the inverse problem multimodal, with discrete families of solutions that fit comparably well but imply very different physical parameters.

The quiescent emission offers a complementary constraint. Between eruptions, QPE hosts show the thermal spectrum of a compact accretion disk, and recent models of such TDE disks constrain the mass and spin of the SMBH and the properties of the disk \citep[e.g.][]{Mummery23,Mummery25b,Guolo25c}. Moreover, TDE disks are generically expected to be warped as a result of their misalignment with the SMBH spin \citep[e.g.][]{Kaaz23,Mummery26b}. Both the SED and the collision timings should show signatures of this effect (as also recently pointed out in \citealt{Chen26}), and a self-consistent solution for both would be strong evidence for the EMRI model and for warped TDE disks, and a powerful probe of the SMBH, EMRI, and disk.

In this paper, we present and apply a framework that exploits this by jointly fitting QPE spectral-timing data with an analytic collision-timing model built on Kerr geodesic frequencies and a self-irradiated warped-disk SED model \citep{Mummery26b} that share a single set of physical parameters. We then fit two sources with relatively regular timings and good SED data, eRO-QPE2 and GSN~069. Section~\ref{sec:methods} describes the data and the two model channels, Section~\ref{sec:results} presents the fits, and Section~\ref{sec:discussion_conclusion} discusses the implications.

\section{Methods} \label{sec:methods}

\subsection{Observations and data reduction} \label{subsec:data}
\textbf{eRO-QPE2:} we analyze the QPE timing data presented in \cite{Arcodia26}, comprising \textit{XMM-Newton} OBSIDs 0941270101--0941270401 (hereafter XMM1--4) taken between June 28 and July 28, 2024. eRO-QPE2 was also observed by \textit{NICER} and \textit{Swift} in brief snapshots between the \textit{XMM} visits. We reduced the \textit{XMM} EPIC-pn data with \texttt{SAS} following standard procedures. Event lists were reprocessed with \texttt{epproc}, screened for background flares by thresholding the 10--12\,keV single-event light curve, and barycenter corrected with the DE440 ephemeris. Source products were extracted from a $33''$ circular aperture with background taken from a nearby source-free region, and background-subtracted 0.3--2\,keV light curves were produced with \texttt{epiclccorr}. Eruption arrival times were measured by fitting each eruption profile with an asymmetric Gaussian, with independent rise and decline widths; the reported times correspond to the fitted peak, with statistical uncertainties from the fit covariance ($\sim$25--60\,s for \textit{XMM} eruptions). \citet{Arcodia26} showed that the timings are insensitive to the choice of eruption model. For \textit{NICER} and \textit{Swift} we adopt the arrival times and uncertainties of \citet{Arcodia26}; their larger errors make them much less constraining than \textit{XMM}. We adopt the UV/X-ray SED data presented in \cite{Wevers25}.

\textbf{GSN~069:} we analyze the \textit{XMM-Newton} and \textit{Chandra} observations presented in \cite{Miniutti25}. For \textit{XMM-Newton} we follow the same reduction procedures outlined above. For the \textit{Chandra} observation (OBSID~22096 on Feb. 14, 2019) we reprocessed the Level~1 event list with \texttt{chandra\_repro} in \textsc{ciao}~4.16 \citep{ciao} following the standard analysis threads. Source events were extracted from the Level~2 event files with a circular aperture centered on the source with background from a nearby source-free region, and the arrival times were corrected to the solar-system barycenter with \texttt{axbary}. We produced a background-subtracted 0.3--2\,keV light curve binned at 100\,s and measured the eruption arrival times with the same asymmetric-Gaussian profile fit as for the \textit{XMM} eruptions. In total, the data contain 15 eruptions across three \textit{XMM-Newton} epochs and one \textit{Chandra} epoch over a 160-day baseline. We adopt the UV/X-ray SED data presented in \citet{Guolo25a}.

\subsection{Warped disk structure}
We model the warped disk with the quasi-steady \citet{Bardeen1975} solution of \citet{Scheuer1996}, in which the annulus at radius $r$ is a ring tilted by $\beta(r)$ and twisted by $\gamma(r)$ (see \citealt{Mummery26b} for details). We first define the coordinate system. 

We work in a right-handed Cartesian frame centered on the black hole,
with the spin axis $\mathbf{\hat{a}}_\bullet = \mathbf{\hat{z}}$ defining the polar direction.  A warped disk occurs in a tidal disruption event when the disrupted star enters on an orbit with angular momentum vector $\mathbf{\hat l}_\infty$ which is misaligned with the black hole's spin axis. 
We orient the remaining axes so that the line of nodes
--- the intersection of the inner and outer disk planes --- lies
along $\mathbf{\hat{y}}$, and the warp peak (maximum $z$-height as $r \to \infty$) lies along $-\mathbf{\hat{x}}$ (this is the original \citealt{Scheuer1996} coordinate convention, which differs from some of the more recent literature, e.g., \citealt{Speicher2025}). The asymptotic
angular-momentum vector $\mathbf{\hat{l}}_\infty$ then lies in the
$xz$-plane 
\begin{equation}
  \mathbf{\hat{l}}_\infty
  = (\sin\theta_{\rm w},0,\;\cos\theta_{\rm w}),
  \label{eq:l_infty}
\end{equation}
so that the misalignment angle between the black hole's and the star's angular momenta is $\theta_{\rm w}$. 

An observer at infinity is specified by two angles: an inclination
$\theta_{\rm obs}$ from the spin axis ($\theta_{\rm obs} = 0^\circ$ is face-on, $\theta_{\rm obs} = 90^\circ$ is
edge-on) and an azimuth $-\phi_{\rm d}$ around it (equivalently, the warp is rotated by $\phi_{\rm d}$ relative to a fixed observer), so that
\begin{equation}
  \mathbf{\hat{o}}
  = (\sin \theta_{\rm obs}\cos\phi_{\rm d},\;
     -\sin \theta_{\rm obs}\sin\phi_{\rm d},\;
     \cos \theta_{\rm obs}).
  \label{eq:observer}
\end{equation}
The convention is that $\phi_{\rm d} = 180^\circ$ places the
observer on the $-\mathbf{\hat{x}}$ side of the disk (i.e., behind the warp direction as $r\to \infty$).

The \cite{Bardeen1975} steady-state solution
\citep{Scheuer1996,Speicher2025} specifies the angular-momentum unit
vector $\mathbf{\hat{l}}(r)$ of each disk annulus at radius $r$. This angular momentum unit vector is the disk surface normal at $r$.   In
our coordinate system,
\begin{align}
  l_x(r) &= \sin(\theta_{\rm w})\,\cos(\gamma(r))\,
             e^{-\gamma(r)},  \label{eq:lx} \\
  l_y(r) &= \sin(\theta_{\rm w})\,\sin(\gamma(r))\,
             e^{-\gamma(r)},  \label{eq:ly} \\
  l_z(r) &= \sqrt{1 - l_x^2 - l_y^2},  \label{eq:lz}
\end{align}
where $\gamma(r) \equiv 2\sqrt{r_{\rm warp}/r}$.  At large radii
($\gamma\to 0$) these reduce to
$\mathbf{\hat{l}} \to (\sin\theta_{\rm w}, 0, \cos\theta_{\rm w})
= \mathbf{\hat{l}}_\infty$, as required. At small radii ($\gamma\to\infty, \mathbf{\hat l}\to \mathbf{\hat z} = \mathbf{\hat a_\bullet}$) the disk lies in the equatorial plane. This is the origin of the global warp. 

Because each annulus is a tilted, twisted ring centered on the origin, both the disk surface and the EMRI crossing condition satisfy 
\begin{equation}
    h = \vec{\mathbf{r}} \cdot \mathbf{\hat l} = 0,
\end{equation}
where $\vec{\mathbf{r}}$ is a vector (either the position on the disk surface or the location of the EMRI) pointing from the origin.

\subsection{Timing model} \label{subsec:timing}

We model the eruption times as the disk crossings of a star following a geodesic around a Kerr SMBH, with one observed eruption per orbit (though we briefly discuss a model with two eruptions per orbit in Section~\ref{subsec:ero2}). The orbiter intersects the disk plane on the polar period $P = 2\pi/\Omega_\theta$, modulated by two relativistic precessions: the apsidal advance of the pericenter at a rate $\dot\omega = \Omega_\theta - \Omega_r$, and the Lense--Thirring precession of the orbital node relative to the disk at a rate $\dot\psi = \Omega_\theta - \Omega_\phi$ (in this notation $\Omega_k$ is the standard relativistic orbital frequency corresponding to the $k$th coordinate). As derived in Appendix~\ref{app:timing_derivation} (see also \citealt{Mummery26c}), the $n$-th eruption time is
\begin{align}
t_n=t_0+nP+\frac{P}{2\pi}&\bigl(E_n-e\sin E_n-f_n\bigr)-\frac{P}{2\pi}\,\chi_n, \label{eq:timing}
\\
\tan\frac{E_n}{2}&=\sqrt{\frac{1-e}{1+e}}\,\tan\frac{f_n}{2}, \\
f_n & =\pi-\omega_n-\chi_n,
\end{align}
where $E_n$ is the classical eccentric anomaly, and the final term results from the different planes of the orbit and disk:
\begin{equation}
    \chi_n = \arctan\!\left[\frac{\sin\beta\,\cos\psi_n}
    {\cos\beta\,\sin i + \sin\beta\,\cos i\,\sin\psi_n}\right],
    \label{eq:chi}
\end{equation}
with $i$ the orbital inclination, $\beta$ the local disk tilt at the crossing radius, and $\omega_n$, $\psi_n$ the pericenter and relative-node phases, advanced at $\dot\omega$ and $\dot\psi$ and evaluated at the crossing time obtained by fixed-point iteration of Eq.~\ref{eq:timing}. For $\beta < i$ the relative node librates and $\chi$ is bounded; for $\beta > i$ it circulates and $\chi$ accumulates $2\pi$ per nodal cycle.

The first correction term in Eq.~\ref{eq:timing} uses the exact Kepler time of flight; for small eccentricity $e$, $\mathcal{M}\equiv E-e\sin E \approx f-2e\sin f+\mathcal{O}(e^2)$, so the offset from uniform angular motion is $\delta t(f)=\frac{P}{2\pi}(E-e\sin E - f) \approx -(eP/\pi)\sin f+\mathcal{O}(e^2)$. With the precessing crossing anomaly $f_n=\pi-\omega_n - \chi_n$ and $\sin(\pi-x)=\sin
x$ this gives 
\begin{equation}
    t_n \approx t_0 + nP
    \;-\; \frac{eP}{\pi}\,\sin\!\left(\omega_n + \chi_n\right)
    \;-\; \frac{P}{2\pi}\,\chi_n,
\end{equation}
showing that for small $e$, the apsidal precession produces a sinusoidal signature in the observed-minus-calculated ($O-C$) timing residuals with a period $N_{\rm aps} \equiv P_{\rm aps}/P = \Omega_\theta / \dot{\omega}$, while longer-term modulations occur (but not strictly sinusoidally) on the nodal timescale, $N_{\rm nod} \equiv P_{\rm nod}/P = \Omega_\theta/|\dot\psi|$.

The three frequencies $(\Omega_r, \Omega_\theta, \Omega_\phi)$ are computed from the Kerr geodesic constants of motion (e.g. \citealt{Schmidt02}), so $P$, $N_{\rm aps}$ and $N_{\rm nod}$ are exact functions of the physical parameters $(M_\bullet, a_\bullet, a, e, i)$, the SMBH mass and spin and the EMRI semi-major axis, eccentricity, and inclination. When the disk is warped, the local tilt entering Eq.~\ref{eq:chi} is evaluated at the instantaneous crossing radius $r_c = a(1-e^2)/[1 - e\cos(\omega_n + \chi_n)]$ from the same warp profile that shapes the SED, coupling the two. The predicted times also include the R{\o}mer and Shapiro delays from the crossing location to the observer. 

For sampling, we reparametrize $(M_\bullet, a_\bullet, a, e) \to (P, P_{\rm aps}, P_{\rm nod}, e)$, as the latter is a more direct representation of the information contained in the timing data. In the weak-field limit,
\begin{equation}
    P \approx \frac{2\pi\, G M_\bullet a^{3/2}}{c^3},\; N_{\rm aps} \approx \frac{a(1-e^2)}{3},\; N_{\rm nod} \approx \frac{[a(1-e^2)]^{3/2}}{2a_\bullet}
\end{equation}
where $a$ is in units of $r_g$. So $N_{\rm aps}$ is set by $a$ and $e$, $N_{\rm nod}$ then by $a_\bullet$, and $P$ by $M_\bullet$. At each likelihood evaluation we invert $(P, P_{\rm aps}, P_{\rm nod}, e)$ to the physical parameters, which in the weak field are approximately
\begin{equation}
    a \approx \frac{3N_{\rm aps}}{1-e^2},\quad a_\bullet \approx \frac{(3N_{\rm aps})^{3/2}}{2N_{\rm nod}},\quad M_\bullet \approx \frac{c^3 P\,(1-e^2)^{3/2}}{2\pi G\,(3N_{\rm aps})^{3/2}}.
    \label{eq:inv}
\end{equation}
These expressions omit the frame-dragging contribution to the apsidal precession, which biases $a_\bullet$ by up to $\approx15\%$ for the high-spin modes, so we use them only as the initial guess for a numerical solve with the exact Kerr frequencies.

This analytic timing model is now implemented in the open-source package \textsc{QPE-FIT},\footnote{\url{https://github.com/joheenc/qpe-fit}} which originally fit the full geodesic EMRI--disk trajectory \citep{Chakraborty25c}. The closed-form model is orders of magnitude faster to evaluate.

The timing likelihood (Section~\ref{subsec:sampling}) includes a systematic floor, $\sigma_{{\rm eff},k}^2 = \sigma_k^2 + (f_{\rm sys,timing} P)^2$, where $f_{\rm sys,timing}$ absorbs stochastic delays between disk crossing and eruption peak from the uncertain emission mechanism. Each observed eruption is matched to the nearest predicted crossing, so no fixed cycle assignment is assumed.

\subsection{SED model} \label{subsec:sed}

The quiescent counterpart of the timing model is the thermal SED of the same disk the orbiter is colliding with. We model a geometrically thin, optically thick disk whose plane transitions from the SMBH equatorial plane at small radii to a misaligned outer plane, with local tilt profile \citep{Mummery26b}
\begin{equation}
    \beta(r) = \arcsin\!\left[\sin\theta_{\rm w}\,
    e^{-2\sqrt{r_{\rm warp}/r}}\right],
    \label{eq:warp}
\end{equation}
where $\theta_{\rm w}$ is the asymptotic outer tilt and $r_{\rm warp}$ the warp radius. The warp azimuth $\phi_{\rm d}$ orients the structure relative to the observer at co-latitude $\theta_{\rm obs}$. This is the same profile, with the same parameters, that sets the local tilt in the timing model (Eq.~\ref{eq:chi}).

The local effective temperature contains a turbulent term and the self-irradiation of the warped surface,
\begin{equation}
    \sigma T^4(r,\phi) = \tfrac{1}{2}\left[Q_{\rm v}(r) + Q_{\rm irr}(r,\phi)\right],
    \; Q_{\rm v} = 2\sigma T_{\rm in}^4 \left(\frac{r}{r_{\rm in}}\right)^{-3},
    \label{eq:temp}
\end{equation}
with a finite-temperature inner boundary condition at the innermost stable circular orbit $r_{\rm in} = r_{\rm ISCO}(a_\bullet)$ and the inner-edge temperature $T_{\rm in}$ a free parameter. A flat disk has $Q_{\rm irr}=0$; in a warped disk, the inner annuli irradiate the outer ones, so $Q_{\rm irr}(r,\phi)$ is computed by summing the received flux over all emitting surface elements with mutual visibility determined by ray tracing, which accounts for any self-shadowing by the warp profile. Because this double surface integral is expensive, we precompute $Q_{\rm irr}$ on a grid in $(\theta_{\rm w}, r_{\rm warp}, a_\bullet)$ and interpolate within it during sampling. The geometry is scale-free in gravitational units, so the grid is independent of SMBH mass. We have verified that the interpolation results in sub-percent errors in the observed spectrum. 

The observed spectrum is the color-corrected multicolor blackbody of the warped surface,
\begin{equation}
    \nu L_\nu = \frac{8\pi h \nu^4}{c^2} \iint
    \frac{\left|\cos\Upsilon_{\rm obs}\right| \, {\rm d}A}
    {f_{\rm c}^4\left[\exp\!\left(h\nu/k_{\rm B} f_{\rm c} T\right) - 1\right]},
    \label{eq:sed}
\end{equation}
where $|\cos\Upsilon_{\rm obs}|\, {\rm d}A$ is the projected area of each surface element toward the observer (with physical surface area ${\rm d}A$) and $f_{\rm c}(T)$ is the color-correction factor of \citet{Done12}. Because the timing may require high spins \citep{Arcodia26}, for which the disk extends close to the horizon, we include the gravitational redshift of the emission. 

The intrinsic spectrum is folded through a single absorbing column $N_{\rm H}$ (Galactic plus host) with the photoelectric cross-sections of \citet{Morrison83} and a \citet{Calzetti00} attenuation law [$E(B-V)$] in the UV, then redshifted to the observed frame. Model configurations in which the warp occults the inner X-ray-emitting region of the disk from the observer's line of sight are rejected, and the observer is restricted to one hemisphere ($\theta_{\rm obs} \le \pi/2$, as $\phi_{\rm d} \in (0, 2\pi)$, this contains the entire observer sky by symmetry).

\begin{table*}
\centering
\caption{Priors of the joint spectral-timing fits for both eRO-QPE2 and GSN~069. We sample these parameters directly. At each likelihood evaluation the periods are inverted to $(M_\bullet, a_\bullet, a)$, and proposals are rejected if $\log_{10} M_\bullet/M_\odot \notin (4,\,8)$, $a_\bullet \notin (0,\,0.998)$, or $a \notin (6,\,800)\,r_g$, or if the warp geometry occults the inner disk from the observer.}
\label{tab:priors}
\begin{tabular}{llll}
\toprule
Parameter & Prior & Unit & Notes \\
\midrule
\multicolumn{4}{l}{\textit{Timing-only parameters}} \\
$i$ & $\mathcal{U}(1,\,179)$ & deg & EMRI orbital inclination \\
$\omega_0$ & $\mathcal{U}(0,\,2\pi)$ & & initial apsidal phase \\
$\psi_0$ & $\mathcal{U}(0,\,2\pi)$ & & initial nodal phase \\
$t_0$ & $\mathcal{U}(-\infty,\,\infty)$ & s & zero-point timing \\
$f_{\rm sys,timing}$ & $\mathcal{U}(10^{-4}, 10^{-1.5})$ &  $P$ & systematic timing error \\
\midrule
\multicolumn{3}{l}{\textit{SED-only parameters}} \\
$T_{\rm in}$ & $\mathcal{U}(5\times10^{4},\,2\times10^{6})$ & K & inner disk temperature \\
$N_{\rm H}$ & log-$\mathcal{U}(10^{21},\,10^{22})$ & cm$^{-2}$ & obscuring column density \\
$E(B-V)$ & $\mathcal{U}(0,\,0.8)$ & mag & reddening \\
$\log_{10} f_{\rm sys,SED}$ & $\mathcal{U}(-3,\,-1)$ & & systematic SED error \\
\midrule
\multicolumn{3}{l}{\textit{Shared parameters}} \\
$P$ & $\mathcal{U}(0,\,\infty)$ & s & EMRI orbital period* \\
$P_{\rm aps}$ & $\mathcal{U}(10P,\,100P)$ & s & apsidal (Schwarzschild) precession period* \\
$P_{\rm nod}$ & $\mathcal{U}(10^{2}P,\,3\times10^{4}P)$ & s & nodal (Lense--Thirring) precession period* \\
$e$ & $\mathcal{U}(0,\,0.6)$ & & orbital eccentricity*\\
$\theta_{\rm w}$ & $\mathcal{U}(0,\,88)$ & deg & asymptotic warp angle\\
$r_{\rm warp}$ & $\mathcal{U}(0.05,\,20)$ & $r_g$ & warp radius \\
$\theta_{\rm obs}$ & $\mathcal{U}(0,\,\pi/2)$ & & observer viewing angle\\
$\phi_{\rm d}$ & $\mathcal{U}(0,\,2\pi)$ & & disk azimuthal orientation\\
\bottomrule
\end{tabular}
\tablenotetext{*}{Although they are orbital parameters, $P$, $P_{\rm aps}$, $P_{\rm nod}$, and $e$ are inverted to $M_\bullet$ and $a_\bullet$ (Eq.~\ref{eq:inv}), which enter the SED model; they are therefore effectively shared.}
\end{table*}

\subsection{Posterior sampling and multimodality} \label{subsec:sampling}

The joint log-likelihood is the sum of the timing and SED terms,
\begin{equation}
    \ln\mathcal{L} = \ln\mathcal{L}_{\rm t} + \ln\mathcal{L}_{\rm SED},
    \label{eq:jointlike}
\end{equation}
each a Gaussian with its own fitted systematic error. The timing term over the $N_{\rm t}$ eruption arrival times $t_k$ (with measurement errors $\sigma_k$) carries an absolute noise floor,
\begin{align}
    \ln\mathcal{L}_{\rm t} &= -\frac{1}{2}\sum_{k=1}^{N_{\rm t}}\Biggl[
    \frac{\left(t_k - t_k^{\rm mod}\right)^2}{\sigma_k^2 + (f_{\rm sys,timing}P)^2} \nonumber\\
    &\qquad + \ln\!\left(2\pi\left[\sigma_k^2 + (f_{\rm sys,timing}P)^2\right]\right)\Biggr]
    \label{eq:liket}
\end{align}
while the SED term over the $N_{\rm s}$ flux points $F_i$ with errors $\varsigma_i$ carries a floor proportional to the model flux,
\begin{align}
    \ln\mathcal{L}_{\rm SED} &= -\frac{1}{2}\sum_{i=1}^{N_{\rm s}}\Biggl[
    \frac{\left(F_i - F_i^{\rm mod}\right)^2}{\varsigma_i^2 + (f_{\rm sys,SED}\,F_i^{\rm mod})^2} \nonumber\\
    &\qquad + \ln\!\left(2\pi\left[\varsigma_i^2 + (f_{\rm sys,SED}\,F_i^{\rm mod})^2\right]\right)\Biggr]
    \label{eq:likesed}
\end{align}
The two systematics play distinct physical roles. The timings are statistically very precise (tens-of-seconds errors on hour-long periods), but the physics linking a disk crossing to the eruption peak is uncertain; $f_{\rm sys,timing}$ absorbs this at up to a few percent of the period and keeps the timing channel from dominating the joint fit. The SED systematic instead absorbs the correlated calibration and scatter from stacking the quiescent spectra, which grows with flux; it is therefore fractional in the model. For eRO-QPE2 we exempt the single UV point from $f_{\rm sys,SED}$, which would otherwise down-weight its unique constraint. For GSN~069, by contrast, with several UV points, the floor applies to both bands. Retaining the normalization terms in Eqs.~\ref{eq:liket}--\ref{eq:likesed} penalizes large systematics, so each settles at the value the data require. The fitted systematics double as goodness-of-fit diagnostics, as a posterior for either systematic that rails against its prior ceiling signals unmodeled additional structure, a behavior we address in Section~\ref{subsec:ero2}.

The joint posterior couples the two data sources through a parameter set involving the disk geometry $(\theta_{\rm w}, r_{\rm warp}, \theta_{\rm obs}, \phi_{\rm d})$ and SMBH parameters ($M_\bullet$, $a_\bullet$), while the orbital elements and phases affect only the timings, and $(T_{\rm in}, N_{\rm H}, E(B-V), f_{\rm sys,SED})$ affect only the SED. For eRO-QPE2 the timings dominate the SMBH constraints, but it is not obvious \textit{a priori} that the timing solutions permit a good SED fit, so the SED provides an independent consistency check. 

We sample the joint posterior with the MCMC ensemble sampler \textsc{emcee} \citep{Foreman-Mackey13} in the $(P, P_{\rm aps}, P_{\rm nod})$ basis. The model has 17 free parameters across the spectral and timing channels, and we list the priors in Table~\ref{tab:priors}. Because the timing likelihood is strongly multimodal in the precession frequencies, the initialization strongly affects the posteriors, so we first locate the modes and then sample each one separately.

The eRO-QPE2 timing data contain 32 eruptions over 30\,d, which admit multiple comparably good solutions with the timing model of Section~\ref{subsec:timing}. Because the eruptions are sampled in a handful of epochs separated by long gaps, the integer cycle count across each gap is ambiguous. Nearby count assignments produce solutions on aliases of the sampling window, each with its own precession timescales and hence its own physical parameters. To find all such solutions systematically, we exploit the fact that the aliases are a property of the observing pattern itself. The spectral window of the eruption sampling (the Fourier transform of the observed cycle numbers) has a sidelobe at every frequency offset $\Delta\nu$ (in inverse orbits) that the gaps between epochs cannot distinguish from zero. As a result, a modulation at frequency $\nu$ is sampled almost identically to one at $\nu+\Delta\nu$. A periodic $O-C$ signal at the apsidal frequency $1/N_{\rm aps}$ therefore has, for every sidelobe, a candidate alias
\begin{equation}
    N_{\rm aps}' = \frac{N_{\rm aps}}{1 + N_{\rm aps}\,\Delta\nu},
    \label{eq:alias}
\end{equation}
at which an alternative solution may exist. We then test every candidate individually, refitting the timing model from many starting phases, and keep those that reproduce the data with a comparable fit statistic.

This procedure yields six distinct, statistically comparable modes, with apsidal super-periods $N_{\rm aps} \simeq 12$--$67$ orbits and nodal super-periods $N_{\rm nod} \simeq 600$--$1400$ orbits. Their timing-only likelihoods span just $\Delta\ln\mathcal{L} \lesssim 7$, with the best modes agreeing within $\Delta\ln\mathcal{L} \simeq 2$. The systematic error is $f_{\rm sys,timing} \simeq 0.8$--$1.0\%$ of the recurrence period ($\sigma_{\rm sys} \approx 65$--$80$\,s; posterior medians) in each mode. The 30\,d baseline therefore constrains the recurrence period precisely ($P = 8055$--$8070$\,s) while leaving the precession timescales ambiguous.

We repeat the same procedure for GSN~069, whose four sparse epochs admit many more near-degenerate timing modes---at least 18, spanning $N_{\rm aps} \simeq 10$--$50$ orbits, all with comparable fit qualities. We then run an independent MCMC sampler per mode to jointly fit the timings and SED and combine them into the reported joint posterior of Section~\ref{subsec:gsn}.

\section{Results} \label{sec:results}
We apply the methods outlined in Section~\ref{sec:methods} to jointly fit the QPE timings and accretion disk SED data of eRO-QPE2 (Section~\ref{subsec:ero2}) and GSN~069 (Section~\ref{subsec:gsn}).

\subsection{eRO-QPE2} \label{subsec:ero2}

\begin{figure*}
    \centering
    \includegraphics[width=\linewidth]{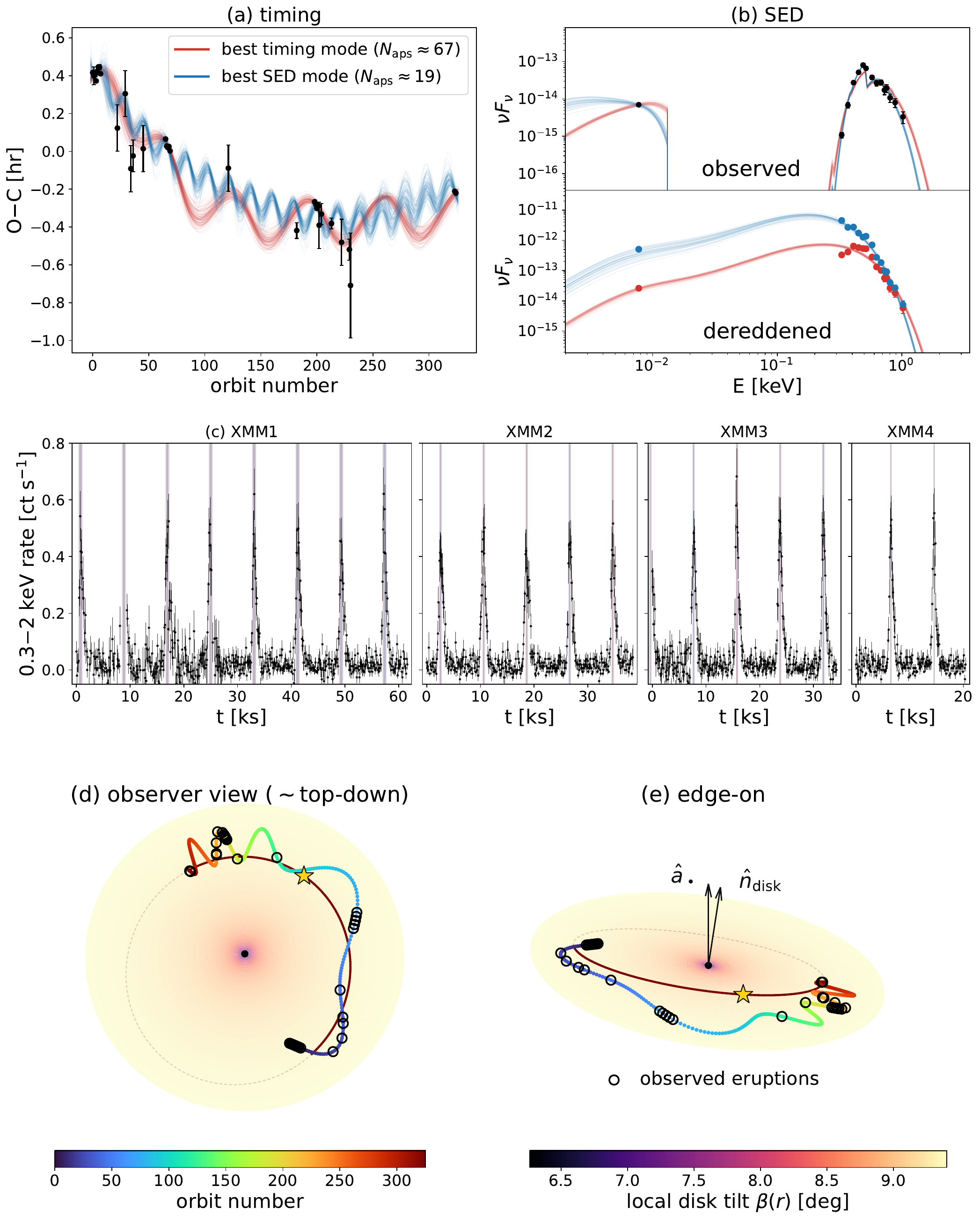}
    \caption{Joint spectral-timing fit of eRO-QPE2 for the best-timing (red) and best-SED (blue) modes. \textbf{(a):} QPE timing $O-C$ data with posterior draws of each model. \textbf{(b):} quiescent broadband SED with the \textit{same} draws, as observed (top) and dereddened (bottom; models with $N_{\rm H}=E(B-V)=0$, data corrected with each mode's fitted absorber). \textbf{(c):} the four \textit{XMM} light curves with the 68\% (dark) and 95\% (light) intervals of the predicted disk-crossing times shaded. \textbf{(d)-(e):} the warped-disk geometry and orbiter--disk crossing locations over the XMM1--4 campaign ($\sim$a quarter of a nodal cycle) for the best-timing mode, seen from the model's observer direction \textbf{(d)} and edge-on \textbf{(e)}.}
    \label{fig:ero2_combined}
\end{figure*}

\begin{figure*}
    \includegraphics[width=\textwidth]{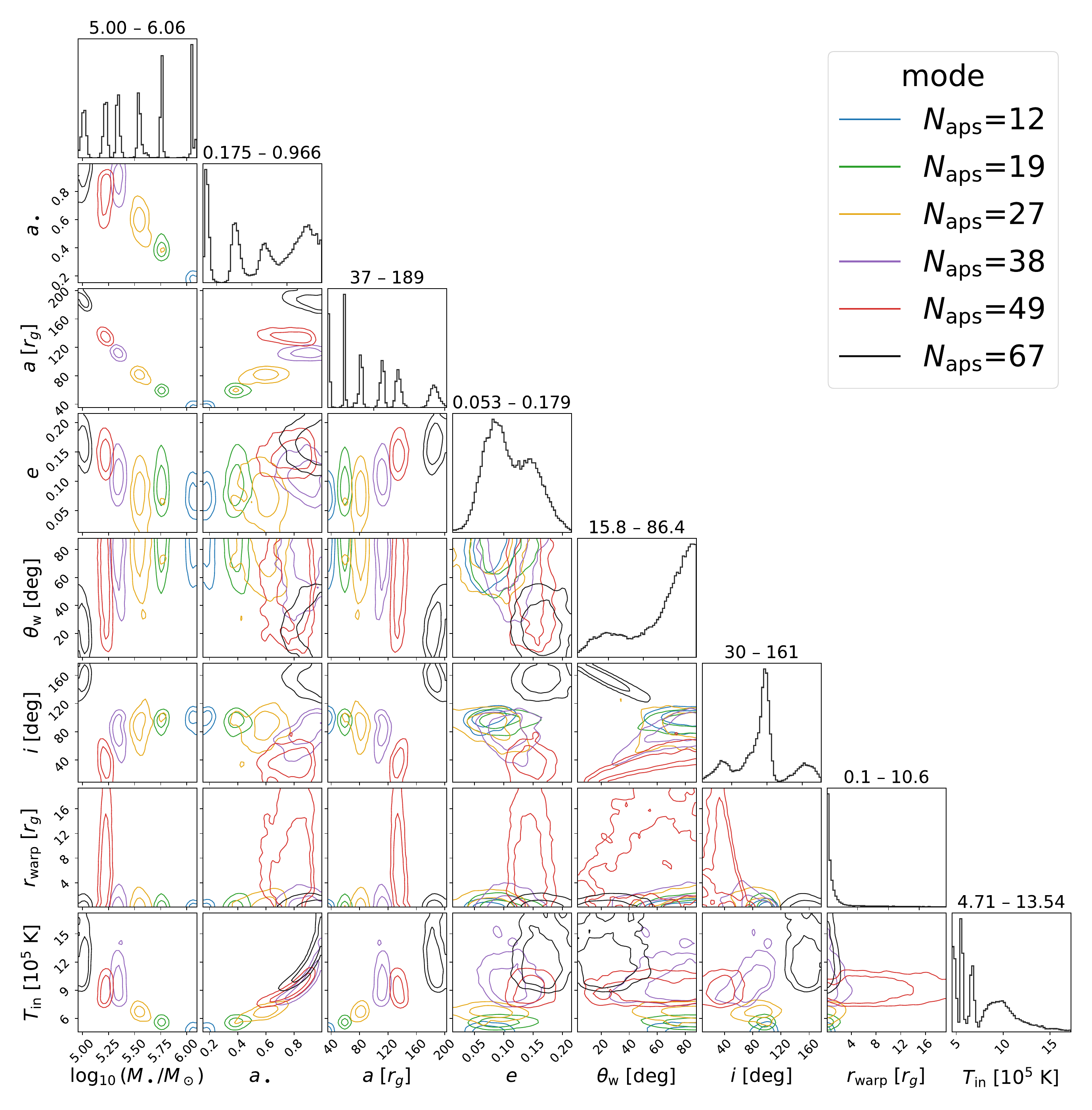}
    \caption{Physical posteriors of the eRO-QPE2 joint spectral-timing fit. The six discrete timing modes (Section~\ref{subsec:sampling}) are shown in different colors, with contours enclosing 68\% and 95\% of the posterior mass, while the diagonal shows the combined marginal distributions. Because the posterior is multimodal we do not quote Gaussian uncertainties; parameter ranges are given in the text. The numbers above each marginal give the 90\% range.}
    \label{fig:ero_corner_physical}
\end{figure*}

\begin{figure}
    \centering
    \includegraphics[width=\linewidth]{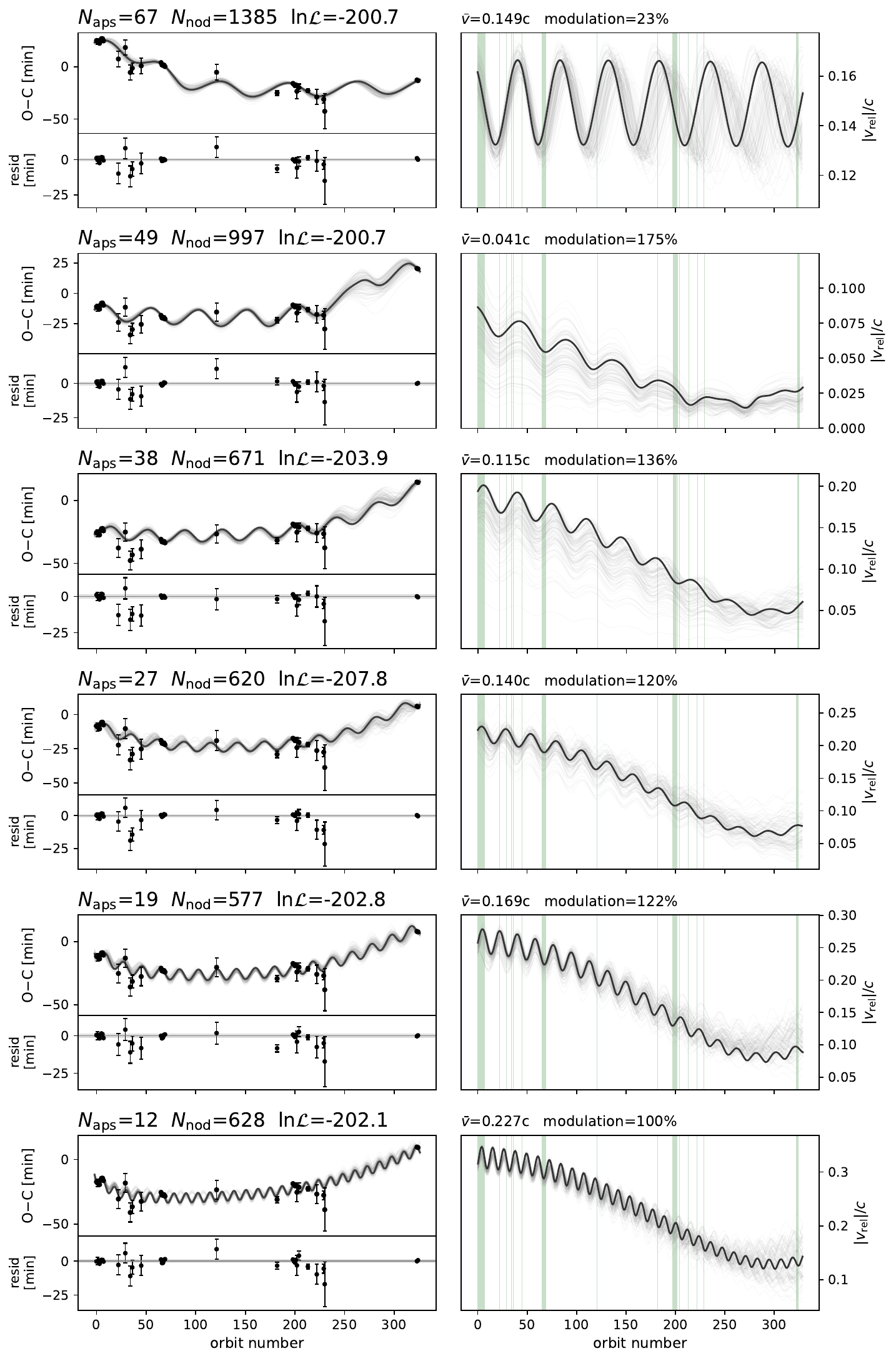}
    \caption{The six timing modes of the eRO-QPE2 XMM1--4 campaign, ordered by decreasing $N_{\rm aps}$. \textbf{Left:} the eruption $O-C$ diagrams and timing residuals, each panel annotated with the mode's timing log-likelihood. \textbf{Right:} predicted EMRI--disk relative velocity at each crossing, with the observing windows shaded. All six fit the timings comparably well (timing $\Delta\ln\mathcal{L} \lesssim 7$), but only the first (largest $N_{\rm aps}$) predicts collision kinematics stable enough to be compatible with the near-constant burst luminosities.}
    \label{fig:ero2_modes}
\end{figure}

\begin{figure}
    \centering
    \includegraphics[width=\linewidth]{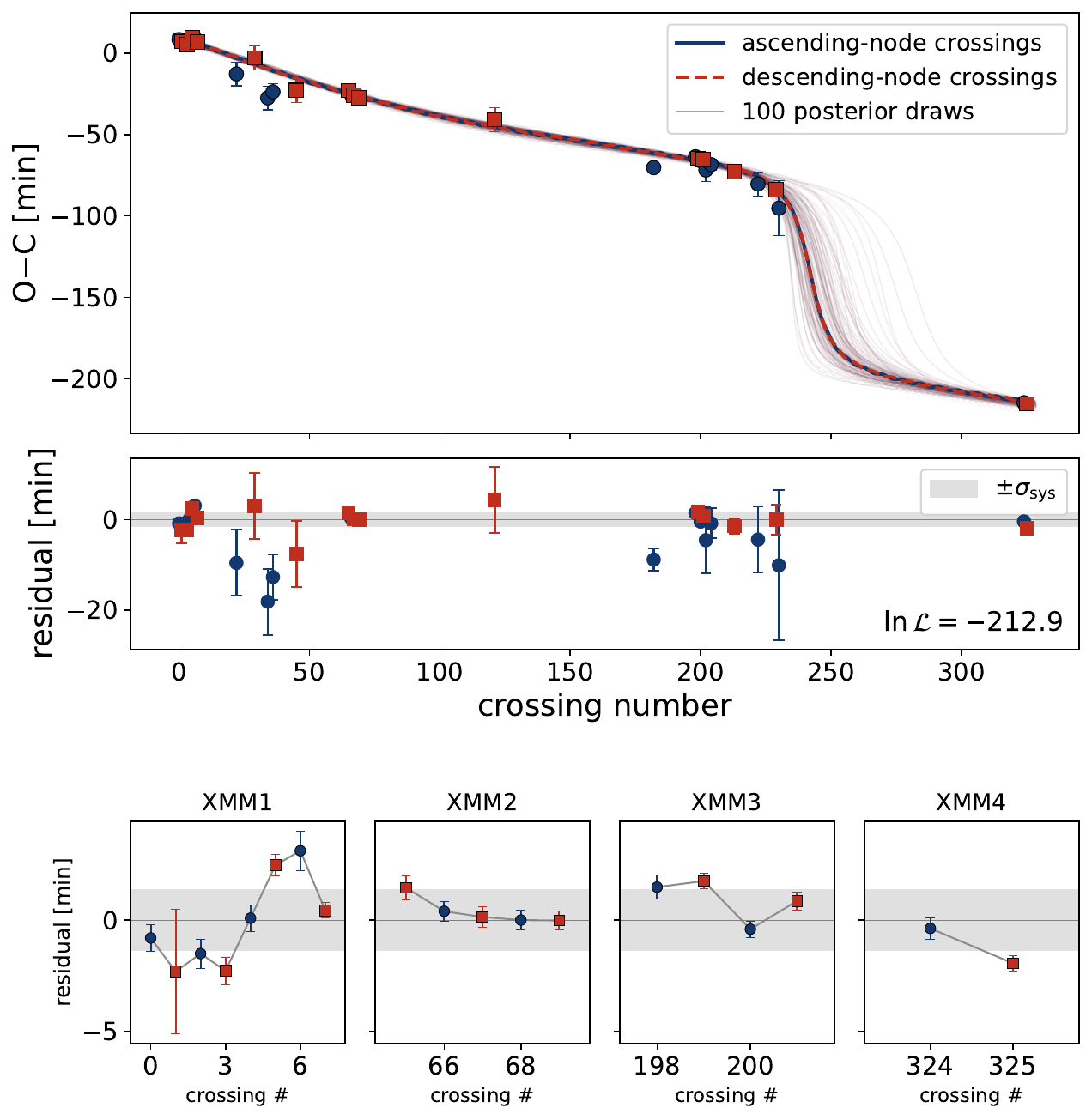}
    \caption{The best-fit model for eRO-QPE2 timings with two observed eruptions per orbit. \textit{Top:} $O-C$ diagram separating eruptions by parity (blue/red for passages through ascending/descending node). \textit{Bottom:} timing residuals colored by parity. The model underfits the data ($\chi^2/{\rm dof} = 44/21$) relative to the model with one observed eruption per orbit, with coherent residual structure visible in the \textit{XMM} points.}
    \label{fig:ero2_twoperorbit}
\end{figure}

Rather than single out one mode, Fig.~\ref{fig:ero2_combined} shows the best-timing ($N_{\rm aps}\approx 67$) and best-SED ($N_{\rm aps}\approx 19$) modes. Across the six modes (Fig.~\ref{fig:ero_corner_physical}), the orbiter has semi-major axis $a\approx40$--$200\,r_g$ and eccentricity $e\approx0.1$, and is inclined by $i\approx 30$--$170^\circ$ to the black-hole spin axis (where values $>90^\circ$ correspond to retrograde orbits). The central SMBH is consistent with $\log_{10} M_\bullet/M_\odot\approx 5$--$6$ and a broad range of spins, $a_\bullet\approx 0.2$--$0.99$. The data prefer a misaligned disk, with asymptotic tilt $\theta_{\rm w} \approx 10$--$90^\circ$, but with a small $r_{\rm warp}\lesssim 1\,r_g$ indicating no evidence of radial structure in the misalignment, though with the modest eccentricities ($e\approx0.1$) implied by our fits the timings would not be sensitive to such a signature. The timings converge to a systematic error of $f_{\rm sys,timing} \approx 0.6$--$1.7\%$ of the recurrence period, consistent with a modest stochastic delay in the crossing-to-eruption radiation processes, and the SED is reproduced across the UV and X-ray bands with an inner-disk temperature $T_{\rm in} \approx 10^{5\text{--}6}$ K. Both modes in Fig.~\ref{fig:ero2_combined} reproduce the observed SED, but their dereddened spectra differ by an order of magnitude (panel b, bottom). With only the absorbed X-ray tail and a single UV point, the absorption and the intrinsic disk luminosity are degenerate, and the more massive SMBH of the best-SED mode implies a brighter disk behind more absorption ($E(B-V)\approx0.46$) compared to the $N_{\rm aps}\approx67$ mode ($E(B-V)\approx0.14$). Multi-band UV photometry or UV spectroscopy, which would measure the slope of the UV continuum and hence the reddening, could break this degeneracy. The full posteriors in the sampled parameters are shown in Appendix~\ref{appendix}.

Figure~\ref{fig:ero2_modes} shows the $O-C$ diagrams of the six modes, which correspond to quite different physical solutions, together with the EMRI--disk relative velocity $v_{\rm rel}$ at each crossing, which sets the shock velocity and hence the eruption properties. Only the $N_{\rm aps}\approx67$ mode keeps $v_{\rm rel}$ nearly constant across the observed eruptions (Section~\ref{sec:discussion_conclusion}), because its long nodal period ($N_{\rm nod}\approx1385$) limits the Lense--Thirring change in the relative orbit--disk inclination that drives $v_{\rm rel}$.

None of these modes fits the full XMM1--5 data, which extend the baseline by $\sim6$ months: every fit degrades by $\Delta\ln\mathcal{L} \geq 90$ and $f_{\rm sys,timing}$ rails to its 3\% ceiling, the signature of unmodeled structure being absorbed into the noise. Recomputing the aliases on the XMM1--5 baseline yields no mode that fits either. The likely cause is secular evolution of the orbit from gravitational-wave emission and/or hydrodynamic drag \citep[e.g.][]{Linial24a,Arcodia24b}, which our steady-state model omits. Even at the bound $|\dot{P}|\lesssim 2\times 10^{-6}$ of \citet{Arcodia26}, the $O-C$ residuals dephase by of order a period ($\dot{P}n^2/2 \approx 1$) after
\begin{equation}
    t_{\rm dephase} \approx 93\;\mathrm{d}\,\bigg(\frac{P_0}{8000\;\rm s}\bigg)\,\bigg(\frac{\dot{P}}{2\times 10^{-6}}\bigg)^{-1/2},
\end{equation}
so only baselines $\ll93$\,d can neglect this evolution. Disk evolution, which we also neglect and which may be as fast or faster, would tighten this further: the quiescent flux evolution of many QPE sources already indicates non-stationary disks on these timescales \citep[e.g.][]{Miniutti23a,Arcodia24a,Nicholl24,Baldini26}, and the QPEs themselves may affect the disk \citep{Linial24c}. This motivates a time-dependent extension of the framework, e.g. with alignment or precession of the outer warp and/or evolution of the disk surface density.

We also tested the alternative hypothesis that two eruptions are produced per orbit, at both the ascending and descending node crossings, with $P_{\rm orb} \approx 16$\,ks. Apsidal motion in this picture generically produces an anti-correlation between the two crossing parities unless the orbit is circular, so the absence of an alternating odd/even sinusoid in the $O-C$ restricts the eccentricity to $e \lesssim 0.05$, as outlined in \citet{Arcodia26}. With the apsidal channel suppressed, the model must produce the observed $O-C$ structure with its nodal term alone. The best such model is shown in Fig.~\ref{fig:ero2_twoperorbit}. It falls short of the models with one observable QPE per orbit by $\Delta\ln\mathcal{L} \approx 12$ ($\sim$$2\times10^{5}:1$ as a likelihood ratio), as the data genuinely contain short-term structure suggesting a rapid oscillatory term. We thus disfavor this interpretation for eRO-QPE2, deferring to future work the open puzzle of what causes the transient long-short alternation in the first observed epoch in 2020 \citep{Arcodia21}.

\subsection{GSN~069} \label{subsec:gsn}

\begin{figure*}
    \centering
    \includegraphics[width=\linewidth]{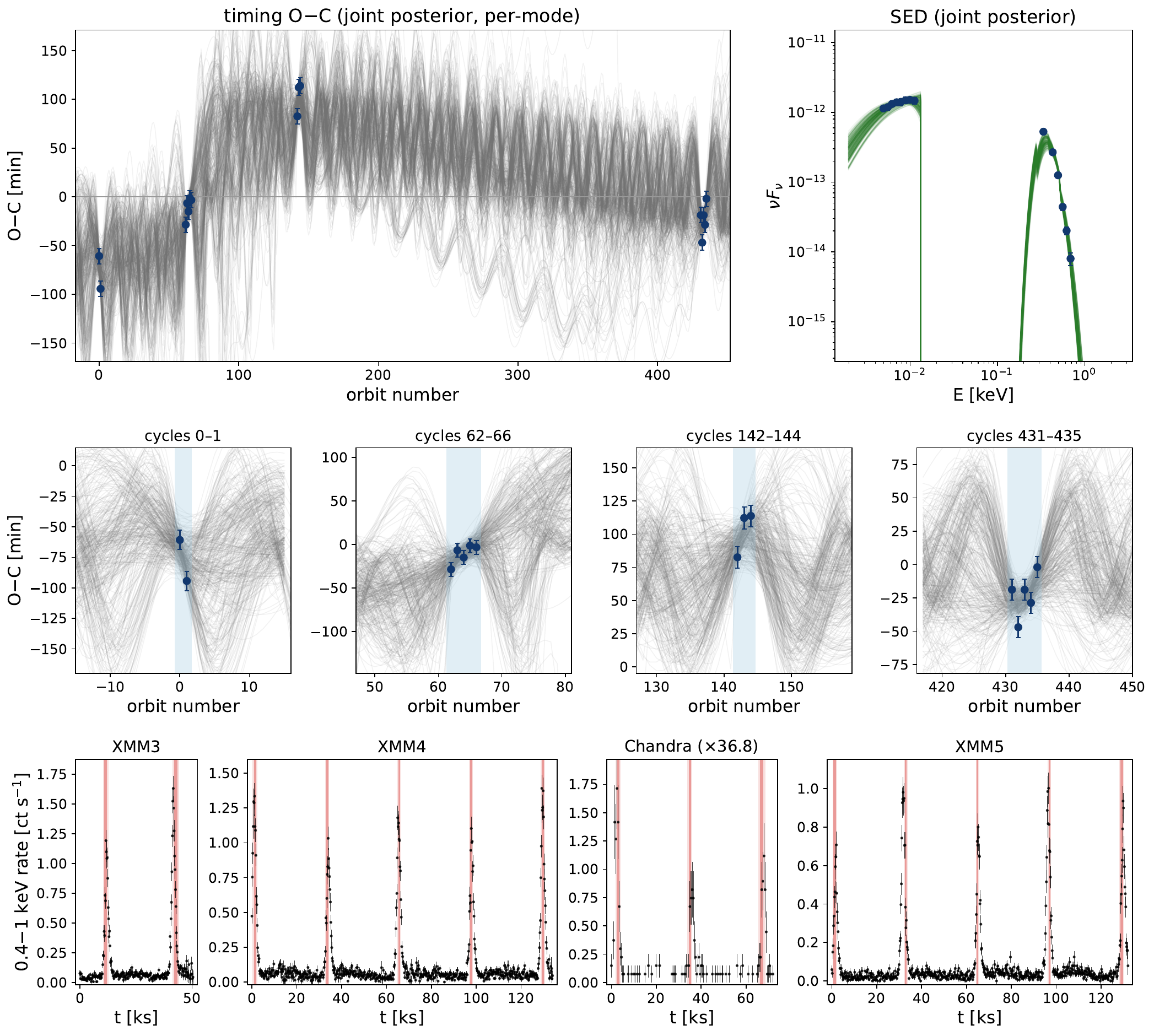}
    \caption{Joint spectral-timing fit of GSN~069. \textit{Top left:} eruption timing $O-C$ residuals with posterior draws from each of the 13 retained modes (within $\Delta\ln\mathcal{L}<10$ of each mode's maximum), each shown against its own linear ephemeris and anchored to the observed eruptions, so all modes pass through the data and fan out across the gaps. The rapid modulation is the apsidal oscillation ($N_{\rm aps}\approx12$--$32$ orbits). \textit{Top right:} quiescent SED with the warped-disk model evaluated for the same posterior draws as the timing panels. \textit{Middle row:} the same trajectories zoomed on each observed epoch: the posterior pinches at the data and fans out in the gaps, reflecting the sparse sampling. \textit{Bottom row:} the 0.4--1\,keV light curves of the four epochs (\textit{XMM-Newton} and \textit{Chandra}, the latter scaled by $\times 36.8$ to the \textit{XMM}-equivalent count rate; panel widths proportional to exposure) with the 68\% (dark) and 95\% (light) intervals of the posterior-predicted eruption times shaded. All panels are generated from the same joint posterior draws.}
    \label{fig:gsn_combined}
\end{figure*}

\begin{figure*}
    \includegraphics[width=\textwidth]{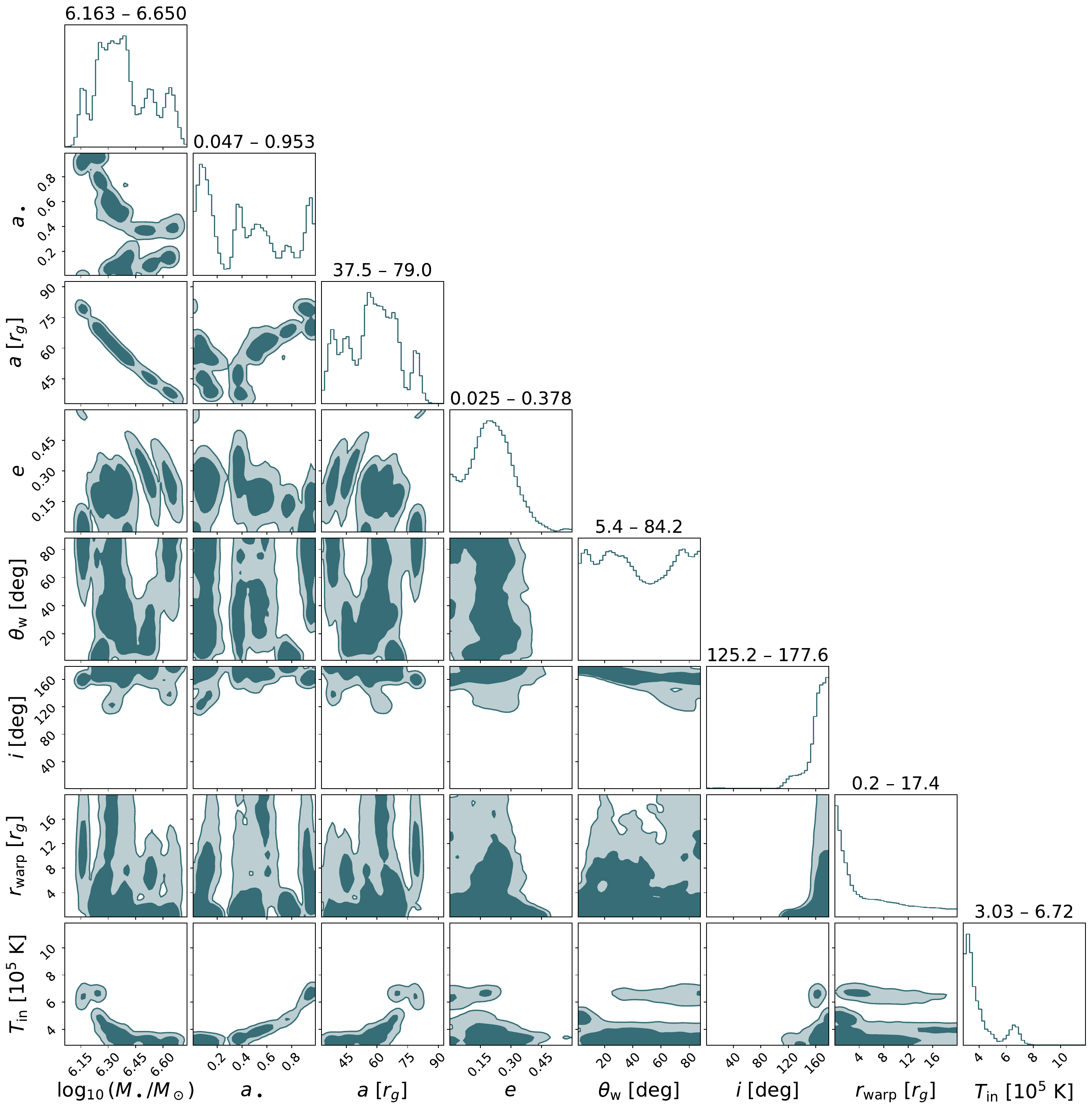}
    \caption{Physical posteriors of the GSN~069 joint spectral-timing fit. The timing modes retained by the SED ($N_{\rm aps}\lesssim 36$; Section~\ref{subsec:gsn}) are combined into a single posterior; contours enclose 68\% and 95\% of the posterior mass. The combined distribution is multimodal, so we quote parameter ranges rather than Gaussian uncertainties in the text. The numbers above each marginal give the 90\% range.}
    \label{fig:gsn_corner_physical}
\end{figure*}

GSN~069 illustrates the even sparser timing regime: with 15 eruptions in four epochs over 160\,d ($\sim$430 orbits), the timings barely constrain the precession frequencies at any timescale. Unlike eRO-QPE2, where the data are well sampled enough to enumerate a handful of modes, GSN~069 admits many closely spaced modes forming a near-continuum of solutions (Fig.~\ref{fig:gsn_combined}), though we note there is still some evidence for disfavored regions in $M_\bullet$ and $a$ among the closely spaced modes (Fig.~\ref{fig:gsn_corner_physical}). The SED disfavors those with $N_{\rm aps}\gtrsim40$, as they require smaller $M_\bullet$ than admitted by the data. We combine the remaining 13 modes ($N_{\rm aps}\lesssim36$) with equal weight. The recurrence period is again tightly constrained ($P\approx31{,}660$--$31{,}750$\,s), but most other parameters are not: $N_{\rm aps}\approx 12$--$32$ orbits, $N_{\rm nod}\approx 300$--$4000$, $\log_{10} M_\bullet/M_\odot \approx 6.16$--$6.65$, $a\approx 37$--$80\,r_g$, $e\approx0.03$--$0.4$, $i\approx 125$--$178^\circ$ (favoring a retrograde orbiter), $r_{\rm warp}\approx0.2$--$17\,r_g$, and $T_{\rm in}\approx 10^{5.5\text{--}5.8}$ K. Moreover, $a_\bullet$ and $\theta_{\rm w}$ are almost entirely unconstrained within the priors. We plot the posterior distributions across the retained modes in Fig.~\ref{fig:gsn_corner_physical}, and further show the full timing and SED parameter posteriors in Appendix~\ref{appendix}.

However, given the more complete UV spectral coverage, the SED constraint contributes appreciably to the posteriors, as a timing-only fit broadens the $M_\bullet$ and $a$ posteriors by 34\% and 39\%. Prior work has found in GSN~069 that SED fitting alone constrains the SMBH mass to $\log_{10} (M_\bullet/M_\odot)\approx 6.4$--$6.7$ (\citealt{Guolo25a,Guolo25b}; the minor discrepancy is because we include self-irradiation by the warped disk, which increases the UV flux at a given SMBH mass). This mass constraint would alone limit the semi-major axis to $\approx 35$--$55\,r_g$. Even within this range there is a continuum of modes, but without the restriction the continuum broadens considerably.

This is the main advantage of joint fitting: for very sparsely sampled QPE sources, the disk SED fit confines the space within which the timing solutions can be identified. This is particularly useful for long-period QPEs, which require larger observing time investments for the same number of orbits. Our results motivate applying the joint spectral-timing framework to the longest-period sources, as well as those with only a few eruption epochs, where the timings alone cannot constrain the apsidal motion.

\section{Discussion and conclusion} \label{sec:discussion_conclusion}

We have presented a joint spectral-timing framework for QPEs driven by orbiter--disk collisions, combining an analytic timing model based on Kerr geodesic frequencies (Section~\ref{subsec:timing}) with a warped-disk SED model (Section~\ref{subsec:sed}) through a shared set of physical parameters. It builds on recent progress in models of warped disks and self-irradiation effects in their SEDs \citep{Mummery26b} and QPE timing models \citep{Chakraborty25c}.

Taken at face value, the eRO-QPE2 solutions describe a $\sim\!10^{5\text{--}6}\,M_\odot$ black hole with a stellar-mass orbiter at $\sim40$--$200\,r_g$ precessing apsidally on timescales of $\sim1$--$6.5$\,d and nodally on $\sim50$--$130$\,d, and colliding with a tilted disk misaligned with the SMBH spin axis by $\sim10$--$90^\circ$. The spectral-timing fit simultaneously reproduces the QPE timings and the quiescent SED (Fig.~\ref{fig:ero2_combined}), albeit with a degenerate family of solutions (Fig.~\ref{fig:ero2_modes}). The constraints come mainly from the timings, as eRO-QPE2 is among the best-sampled QPEs \citep{Arcodia26}, but notably the SED can independently be reproduced within each timing solution.

Although the six eRO-QPE2 solutions fit the timings comparably well, we modestly favor the $N_{\rm aps}\approx67$ solution on physical grounds. In collision models, each eruption is powered either by shocked disk material intercepted by the orbiter, or by shocked stellar debris as it collides with the disk \citep{Linial25}. The flare energy budget thus scales as the kinetic energy of the intercepted mass, $E_{\rm rad} = \eta_{\rm rad} M_{\rm ej}\,v_{\rm rel}^2$, where $\eta_{\rm rad}$ is the radiative efficiency assumed to be $\approx 0.1$ \citep[e.g.][]{Linial25,Huang25,Yao26,Chakraborty26b}. Across the 32 observed eruptions, $v_{\rm rel}$ varies by only a factor of $\approx1.3$ in the $N_{\rm aps}\approx67$ solution, but by factors of $\approx2.5$--$4$ in the other five (Fig.~\ref{fig:ero2_modes}). At fixed $\eta_{\rm rad} M_{\rm ej}$, the latter would imply $\approx7$--$18\times$ changes in eruption energy, which are difficult to reconcile with the remarkably stable luminosities, durations, and duty cycle of the eRO-QPE2 eruptions \citep{Arcodia24b,Arcodia26}. We note, however, that depending on the nature of the orbiter and the emission process, both the shocked mass and the radiative efficiency may strongly depend on $v_{\rm rel}$. Given the uncertainties in the emission mechanism, this is a modest rather than definitive preference, but it shows how eruption energetics can break degeneracies that the timings alone cannot. Complementarily, timing models as presented here constrain flare properties as a function of $v_{\rm rel}$, and may be used to inform theoretical models of the radiative processes that power QPE flares.

The quiescent UV/optical emission of eRO-QPE2 is dominated by internal (turbulent) heating in all six solutions. Switching off self-irradiation at fixed geometry and $T_{\rm in}$ lowers the model flux by only $\approx1$--$13\%$ in the UV and $\lesssim6\%$ in the optical, so none of the fits is irradiation-dominated: although most solutions are strongly tilted, their small warp radii leave the outer disk nearly coplanar with the inner disk. A strongly warped outer disk would instead intercept much of the central emission and reprocess the eruptions into delayed UV/optical counterparts; the non-detection of UV variability in eRO-QPE2 \citep{Wevers25} thus disfavors such geometries, consistent with our fits.

GSN~069, by contrast, hosts a larger SMBH ($M_\bullet\sim10^{6.2\text{--}6.7}\,M_\odot$), and its much sparser timings admit many physically distinct solutions. It represents the opposite regime, in which the SED bounds the SMBH mass and hence the allowed orbits; together, the two sources illustrate the complementary strengths of joint spectral-timing fits.

Moreover, whereas previous modeling of the QPE timing signatures invoked rigid-body precession of the accretion disk \citep{Franchini23,Chakraborty24,Miniutti25,Zhou24b,Arcodia26}, our fits show that it is not required. An orbit precessing relativistically through a static warped disk, with one observable QPE per orbit, reproduces the observed $O-C$ structure on its own. \citet{Chen26} recently reached a similar conclusion from QPE peak-brightness variations. Such a quasi-steady warp is a more physically natural alternative to a coherently precessing disk over long timescales \citep{Mummery26a}, though even warps should eventually evolve and align with the SMBH spin.

However, given that both current datasets are fundamentally limited in their ability to pick out a unique solution \textit{even within the orbiter-disk collision model}, let alone falsify non-EMRI-based models, it is premature to claim unique evidence for any model or a definitive solution within it. At this stage, any global fit reporting a single unimodal posterior \citep[e.g.][]{Zhou24a,Zhou24b,Zhou25} should thus be treated with caution (see also \citealt{Arcodia26}).

Still, the path forward is promising. It is nontrivial that one SMBH+disk system matches both the quiescent SED and the eruption timings, two independent data channels, and the existing data already carry \textit{some} constraining power: for instance, the SMBH mass posteriors of GSN~069 and eRO-QPE2 are completely distinct. This provides circumstantial support for the orbiter-disk collision scenario, and represents a step towards the long-anticipated goal of EMRI parameter inference with QPE timings \citep[e.g.][]{Xian21,Franchini23,Linial23b,Zhou24b,Chakraborty25c}.

The clearest next step is to obtain denser timing data. Sampling many consecutive (or near-consecutive) eruptions in one source, ideally over at least one full apsidal cycle, is the most direct way to break the multimodality and would be a decisive test of the EMRI model. A good fit to such data would be strong evidence that QPEs arise from EMRIs colliding with (warped) disks, since the predicted modulation on the apsidal and nodal periods is not easily produced by other models, e.g. disk instabilities. Our work thus represents a promising step toward a physical understanding of QPEs and toward realizing their potential as precision SMBH probes.

\section*{Acknowledgments}
A.M. acknowledges support from the Ambrose Monell Foundation, the W.M. Keck Foundation and the John N. Bahcall Fellowship Fund at the Institute for Advanced Study. Support for this work was provided by NASA through the NASA Hubble Fellowship grant \#HST-HF2-51581.001-A awarded by the Space Telescope Science Institute, which is operated by the Association of Universities for Research in Astronomy, Inc., for NASA, under contract NAS5-26555.
\bibliography{refs}{}

@article{Bardeen1975,
  author  = {{Bardeen}, J.~M. and {Petterson}, J.~A.},
  title   = {The {L}ense-{T}hirring Effect and Accretion Disks
             around {K}err Black Holes},
  journal = {\apjl},
  year    = {1975},
  volume  = {195},
  pages   = {L65--L67},
  doi     = {10.1086/181711},
  adsurl  = {https://ui.adsabs.harvard.edu/abs/1975ApJ...195L..65B}
}

@ARTICLE{Arcodia24b,
       author = {{Arcodia}, R. and {Linial}, I. and {Miniutti}, G. and {Franchini}, A. and {Giustini}, M. and {Bonetti}, M. and {Sesana}, A. and {Soria}, R. and {Chakraborty}, J. and {Dotti}, M. and {Kara}, E. and {Merloni}, A. and {Ponti}, G. and {Vincentelli}, F.},
        title = "{Ticking away: The long-term X-ray timing and spectral evolution of eRO-QPE2}",
      journal = {\aap},
         year = 2024,
        month = oct,
       volume = {690},
          eid = {A80},
        pages = {A80},
          doi = {10.1051/0004-6361/202451218},
archivePrefix = {arXiv},
       eprint = {2406.17020},
 primaryClass = {astro-ph.HE},
       adsurl = {https://ui.adsabs.harvard.edu/abs/2024A&A...690A..80A}
}

@article{Speicher2025,
  author  = {{Speicher}, J. and {Blaes}, O.},
  title   = {Thermal Spectra of Warped and Broken Accretion Disks},
  journal = {\apj},
  year    = {2025},
  volume  = {981},
  number  = {1},
  pages   = {61},
  doi     = {10.3847/1538-4357/adae00},
  eprint  = {2501.14629},
  adsurl  = {https://ui.adsabs.harvard.edu/abs/2025ApJ...981...61S}
}

@article{Scheuer1996,
  author  = {{Scheuer}, P.~A.~G. and {Feiler}, R.},
  title   = {The realignment of a black hole misaligned with its
             accretion disc},
  journal = {\mnras},
  year    = {1996},
  volume  = {282},
  number  = {1},
  pages   = {291--294},
  doi     = {10.1093/mnras/282.1.291},
  adsurl  = {https://ui.adsabs.harvard.edu/abs/1996MNRAS.282..291S}
}

@ARTICLE{Arcodia21,
       author = {{Arcodia}, R. and {Merloni}, A. and {Nandra}, K. and {Buchner}, J. and {Salvato}, M. and {Pasham}, D. and {Remillard}, R. and {Comparat}, J. and {Lamer}, G. and {Ponti}, G. and {Malyali}, A. and {Wolf}, J. and {Arzoumanian}, Z. and {Bogensberger}, D. and {Buckley}, D.~A.~H. and {Gendreau}, K. and {Gromadzki}, M. and {Kara}, E. and {Krumpe}, M. and {Markwardt}, C. and {Ramos-Ceja}, M.~E. and {Rau}, A. and {Schramm}, M. and {Schwope}, A.},
        title = "{X-ray quasi-periodic eruptions from two previously quiescent galaxies}",
      journal = {\nat},
         year = 2021,
        month = apr,
       volume = {592},
       number = {7856},
        pages = {704-707},
          doi = {10.1038/s41586-021-03394-6},
archivePrefix = {arXiv},
       eprint = {2104.13388},
 primaryClass = {astro-ph.HE},
       adsurl = {https://ui.adsabs.harvard.edu/abs/2021Natur.592..704A}
}

@ARTICLE{Arcodia26,
       author = {{Arcodia}, R. and {Miniutti}, G. and {Chakraborty}, J. and {Franchini}, A. and {Giustini}, M. and {Linial}, I. and {Mummery}, A. and {Bertassi}, L. and {Bonetti}, M. and {Kara}, E. and {Merloni}, A. and {Motta}, A. and {Ponti}, G. and {Quintin}, E. and {Soria}, R. and {Baldini}, P. and {Buchner}, J. and {Dotti}, M. and {Fragile}, P.~C. and {Ingram}, A. and {Middleton}, M. and {Panagiotou}, C. and {Sesana}, A. and {Yao}, P. and {Rau}, A. and {Vincentelli}, F.~M. and {Guolo}, M. and {Saxton}, R.},
        title = "{Even a Precessing Clock Is Right Twice per Orbit{\textemdash}The Superperiods of eRO-QPE2 and Challenges for Quasiperiodic Eruption Orbital Models}",
      journal = {\apj},
         year = 2026,
        month = jun,
       volume = {1003},
       number = {2},
          eid = {148},
        pages = {148},
          doi = {10.3847/1538-4357/ae6078},
archivePrefix = {arXiv},
       eprint = {2604.09788},
 primaryClass = {astro-ph.HE},
       adsurl = {https://ui.adsabs.harvard.edu/abs/2026ApJ..1003..148A}
}

@ARTICLE{Hernandez25a,
       author = {{Hern{\'a}ndez-Garc{\'\i}a}, Lorena and {Chakraborty}, Joheen and {S{\'a}nchez-S{\'a}ez}, Paula and {Ricci}, Claudio and {Cuadra}, Jorge and {McKernan}, Barry and {Ford}, K.~E. Saavik and {Ar{\'e}valo}, Patricia and {Rau}, Arne and {Arcodia}, Riccardo and {Kara}, Erin and {Liu}, Zhu and {Merloni}, Andrea and {Bruni}, Gabriele and {Goodwin}, Adelle and {Arzoumanian}, Zaven and {Assef}, Roberto J. and {Baldini}, Pietro and {Bayo}, Amelia and {Bauer}, Franz E. and {Bernal}, Santiago and {Brightman}, Murray and {Calistro Rivera}, Gabriela and {Gendreau}, Keith and {Homan}, David and {Krumpe}, Mirko and {Lira}, Paulina and {Mart{\'\i}nez-Aldama}, Mary Loli and {Salvato}, Mara and {Sotomayor}, Bel{\'e}n},
        title = "{Discovery of extreme quasi-periodic eruptions in a newly accreting massive black hole}",
      journal = {Nature Astronomy},
         year = 2025,
        month = jun,
       volume = {9},
        pages = {895-906},
          doi = {10.1038/s41550-025-02523-9},
archivePrefix = {arXiv},
       eprint = {2504.07169},
 primaryClass = {astro-ph.HE},
       adsurl = {https://ui.adsabs.harvard.edu/abs/2025NatAs...9..895H}
}

@ARTICLE{Chakraborty25a,
       author = {{Chakraborty}, Joheen and {Kara}, Erin and {Arcodia}, Riccardo and {Buchner}, Johannes and {Giustini}, Margherita and {Hern{\'a}ndez-Garc{\'\i}a}, Lorena and {Linial}, Itai and {Masterson}, Megan and {Miniutti}, Giovanni and {Mummery}, Andrew and {Panagiotou}, Christos and {Quintin}, Erwan and {S{\'a}nchez-S{\'a}ez}, Paula},
        title = "{Discovery of Quasiperiodic Eruptions in the Tidal Disruption Event and Extreme Coronal Line Emitter AT2022upj: Implications for the QPE/TDE Fraction and a Connection to ECLEs}",
      journal = {\apjl},
         year = 2025,
        month = apr,
       volume = {983},
       number = {2},
          eid = {L39},
        pages = {L39},
          doi = {10.3847/2041-8213/adc2f8},
archivePrefix = {arXiv},
       eprint = {2503.19013},
 primaryClass = {astro-ph.HE},
       adsurl = {https://ui.adsabs.harvard.edu/abs/2025ApJ...983L..39C}
}

@ARTICLE{Chakraborty26,
       author = {{Chakraborty}, Joheen and {Rappaport}, Saul A. and {Arcodia}, Riccardo and {Linial}, Itai and {Miniutti}, Giovanni and {Burdge}, Kevin B. and {Cuadra}, Jorge and {Giustini}, Margherita and {Hern{\'a}ndez-Garc{\'\i}a}, Lorena and {Kara}, Erin and {Ricci}, Claudio and {S{\'a}nchez-S{\'a}ez}, Paula and {Yao}, Philippe},
        title = "{A Positive Period Derivative in the Quasiperiodic Eruptions of ZTF19acnskyy}",
      journal = {\apjl},
         year = 2026,
        month = apr,
       volume = {1001},
       number = {1},
          eid = {L6},
        pages = {L6},
          doi = {10.3847/2041-8213/ae548b},
archivePrefix = {arXiv},
       eprint = {2602.16776},
 primaryClass = {astro-ph.HE},
       adsurl = {https://ui.adsabs.harvard.edu/abs/2026ApJ..1001L...6C}
}

@ARTICLE{Chakraborty26b,
       author = {{Chakraborty}, Joheen and {Kara}, Erin and {Lu}, Wenbin and {Metzger}, Brian D. and {Kosec}, Peter and {Arcodia}, Riccardo and {Linial}, Itai and {Aspegren}, Olivia and {Behar}, Ehud and {Bhattacharyya}, Sudip and {Giustini}, Margherita and {Hernandez-Garcia}, Lorena and {Kasen}, Daniel and {Miniutti}, Giovanni and {Paerels}, Frits and {Quintin}, Erwan and {Ricci}, Claudio and {Rogantini}, Daniele and {Sanchez-Saez}, Paula and {Zaidouni}, Fatima},
        title = "{Relativistic outflows power a quasi-periodic eruption: constraints on energetics, mass loss, and emission mechanisms}",
      journal = {arXiv e-prints},
         year = 2026,
        month = aug,
          eid = {arXiv:2608.28507},
        pages = {arXiv:2608.28507},
          doi = {10.48550/arXiv.2608.28507},
archivePrefix = {arXiv},
       eprint = {2608.28507},
 primaryClass = {astro-ph.HE},
       adsurl = {https://ui.adsabs.harvard.edu/abs/2026arXiv260828507C}
}

@ARTICLE{Chakraborty25c,
       author = {{Chakraborty}, Joheen and {Drummond}, Lisa V. and {Bonetti}, Matteo and {Franchini}, Alessia and {Kejriwal}, Shubham and {Miniutti}, Giovanni and {Arcodia}, Riccardo and {Hughes}, Scott A. and {Duque}, Francisco and {Kara}, Erin and {Sesana}, Alberto and {Giustini}, Margherita and {Motta}, Amedeo and {Burdge}, Kevin},
        title = "{Prospects for EMRI/MBH Parameter Estimation Using Quasiperiodic Eruption Timings: Short-timescale Analysis}",
      journal = {\apj},
         year = 2025,
        month = oct,
       volume = {992},
       number = {1},
          eid = {120},
        pages = {120},
          doi = {10.3847/1538-4357/ae003b},
archivePrefix = {arXiv},
       eprint = {2508.20162},
 primaryClass = {astro-ph.HE},
       adsurl = {https://ui.adsabs.harvard.edu/abs/2025ApJ...992..120C}
}

@ARTICLE{Arcodia25,
       author = {{Arcodia}, R. and {Baldini}, P. and {Merloni}, A. and {Rau}, A. and {Nandra}, K. and {Chakraborty}, J. and {Goodwin}, A.~J. and {Page}, M.~J. and {Buchner}, J. and {Masterson}, M. and {Monageng}, I. and {Arzoumanian}, Z. and {Buckley}, D. and {Kara}, E. and {Ponti}, G. and {Ramos-Ceja}, M.~E. and {Salvato}, M. and {Gendreau}, K. and {Grotova}, I. and {Krumpe}, M.},
        title = "{SRG/eROSITA No. 5: Discovery of Quasiperiodic Eruptions Every {\ensuremath{\sim}}3.7 days from a Galaxy at z > 0.1}",
      journal = {\apj},
         year = 2025,
        month = aug,
       volume = {989},
       number = {1},
          eid = {13},
        pages = {13},
          doi = {10.3847/1538-4357/adec9b},
archivePrefix = {arXiv},
       eprint = {2506.17138},
 primaryClass = {astro-ph.HE},
       adsurl = {https://ui.adsabs.harvard.edu/abs/2025ApJ...989...13A}
}

@ARTICLE{Arcodia24a,
       author = {{Arcodia}, R. and {Liu}, Z. and {Merloni}, A. and {Malyali}, A. and {Rau}, A. and {Chakraborty}, J. and {Goodwin}, A. and {Buckley}, D. and {Brink}, J. and {Gromadzki}, M. and {Arzoumanian}, Z. and {Buchner}, J. and {Kara}, E. and {Nandra}, K. and {Ponti}, G. and {Salvato}, M. and {Anderson}, G. and {Baldini}, P. and {Grotova}, I. and {Krumpe}, M. and {Maitra}, C. and {Miller-Jones}, J.~C.~A. and {Ramos-Ceja}, M.~E.},
        title = "{The more the merrier: SRG/eROSITA discovers two further galaxies showing X-ray quasi-periodic eruptions}",
      journal = {\aap},
         year = 2024,
        month = apr,
       volume = {684},
          eid = {A64},
        pages = {A64},
          doi = {10.1051/0004-6361/202348881},
archivePrefix = {arXiv},
       eprint = {2401.17275},
 primaryClass = {astro-ph.HE},
       adsurl = {https://ui.adsabs.harvard.edu/abs/2024A&A...684A..64A}
}

@ARTICLE{Morrison83,
       author = {{Morrison}, R. and {McCammon}, D.},
        title = "{Interstellar photoelectric absorption cross sections, 0.03-10 keV.}",
      journal = {\apj},
         year = 1983,
        month = jul,
       volume = {270},
        pages = {119-122},
          doi = {10.1086/161102},
       adsurl = {https://ui.adsabs.harvard.edu/abs/1983ApJ...270..119M}
}

@ARTICLE{Nicholl24,
       author = {{Nicholl}, M. and {Pasham}, D.~R. and {Mummery}, A. and {Guolo}, M. and {Gendreau}, K. and {Dewangan}, G.~C. and {Ferrara}, E.~C. and {Remillard}, R. and {Bonnerot}, C. and {Chakraborty}, J. and {Hajela}, A. and {Dhillon}, V.~S. and {Gillan}, A.~F. and {Greenwood}, J. and {Huber}, M.~E. and {Janiuk}, A. and {Salvesen}, G. and {van Velzen}, S. and {Aamer}, A. and {Alexander}, K.~D. and {Angus}, C.~R. and {Arzoumanian}, Z. and {Auchettl}, K. and {Berger}, E. and {de Boer}, T. and {Cendes}, Y. and {Chambers}, K.~C. and {Chen}, T. -W. and {Chornock}, R. and {Fulton}, M.~D. and {Gao}, H. and {Gillanders}, J.~H. and {Gomez}, S. and {Gompertz}, B.~P. and {Fabian}, A.~C. and {Herman}, J. and {Ingram}, A. and {Kara}, E. and {Laskar}, T. and {Lawrence}, A. and {Lin}, C. -C. and {Lowe}, T.~B. and {Magnier}, E.~A. and {Margutti}, R. and {McGee}, S.~L. and {Minguez}, P. and {Moore}, T. and {Nathan}, E. and {Oates}, S.~R. and {Patra}, K.~C. and {Ramsden}, P. and {Ravi}, V. and {Ridley}, E.~J. and {Sheng}, X. and {Smartt}, S.~J. and {Smith}, K.~W. and {Srivastav}, S. and {Stein}, R. and {Stevance}, H.~F. and {Turner}, S.~G.~D. and {Wainscoat}, R.~J. and {Weston}, J. and {Wevers}, T. and {Young}, D.~R.},
        title = "{Quasi-periodic X-ray eruptions years after a nearby tidal disruption event}",
      journal = {\nat},
         year = 2024,
        month = oct,
       volume = {634},
       number = {8035},
        pages = {804-808},
          doi = {10.1038/s41586-024-08023-6},
archivePrefix = {arXiv},
       eprint = {2409.02181},
 primaryClass = {astro-ph.HE},
       adsurl = {https://ui.adsabs.harvard.edu/abs/2024Natur.634..804N}
}

@ARTICLE{Chakraborty21,
       author = {{Chakraborty}, Joheen and {Kara}, Erin and {Masterson}, Megan and {Giustini}, Margherita and {Miniutti}, Giovanni and {Saxton}, Richard},
        title = "{Possible X-Ray Quasi-periodic Eruptions in a Tidal Disruption Event Candidate}",
      journal = {\apjl},
         year = 2021,
        month = nov,
       volume = {921},
       number = {2},
          eid = {L40},
        pages = {L40},
          doi = {10.3847/2041-8213/ac313b},
archivePrefix = {arXiv},
       eprint = {2110.10786},
 primaryClass = {astro-ph.HE},
       adsurl = {https://ui.adsabs.harvard.edu/abs/2021ApJ...921L..40C}
}

@ARTICLE{Mummery26a,
       author = {{Mummery}, Andrew and {Quataert}, Eliot},
        title = "{The chirping of Lense-Thirring precession in tidal disruption event accretion flows}",
      journal = {arXiv e-prints},
         year = 2026,
        month = sep,
          eid = {arXiv:2609.22559},
        pages = {arXiv:2609.22559},
archivePrefix = {arXiv},
       eprint = {2609.22559},
 primaryClass = {astro-ph.HE},
       adsurl = {https://ui.adsabs.harvard.edu/abs/2026arXiv260922559M}
}

@ARTICLE{Mummery26b,
       author = {{Mummery}, Andrew and {Quataert}, Eliot},
        title = "{Observational signatures of warped accretion flows in tidal disruption events}",
      journal = {arXiv e-prints},
         year = 2026,
        month = sep,
          eid = {arXiv:2609.29787},
        pages = {arXiv:2609.29787},
          doi = {10.48550/arXiv.2609.29787},
archivePrefix = {arXiv},
       eprint = {2609.29787},
 primaryClass = {astro-ph.HE},
       adsurl = {https://ui.adsabs.harvard.edu/abs/2026arXiv260929787M}
}

@ARTICLE{Mummery26c,
       author = {{Mummery}, Andrew and {Chakraborty}, Joheen and {Quataert}, Eliot},
        title = "{Collisions with warped disks}",
      journal = {in prep.},
         year = 2026
}

@ARTICLE{Chen26,
       author = {{Chen}, Bo-An and {You}, Bei and {Miniutti}, Giovanni and {Jiang}, Ning and {Pan}, Zhen and {Yang}, Tao and {Fan}, Xi-Long and {Liao}, Kai and {Ding}, Xu-Heng and {Zhu}, Zong-Hong and {Yang}, Shuai-Kang and {Xu}, Sai-En and {He}, Han and {Fan}, Xiao},
        title = "{QPEs from Warped Disk Collisions with EMRIs: Brightness-Recurrence Diagram and Gravitational-Wave Follow-up}",
      journal = {arXiv e-prints},
         year = 2026,
        month = may,
          eid = {arXiv:2605.24905},
        pages = {arXiv:2605.24905},
          doi = {10.48550/arXiv.2605.24905},
archivePrefix = {arXiv},
       eprint = {2605.24905},
 primaryClass = {astro-ph.HE},
       adsurl = {https://ui.adsabs.harvard.edu/abs/2026arXiv260524905C}
}

@ARTICLE{Franchini23,
       author = {{Franchini}, Alessia and {Bonetti}, Matteo and {Lupi}, Alessandro and {Miniutti}, Giovanni and {Bortolas}, Elisa and {Giustini}, Margherita and {Dotti}, Massimo and {Sesana}, Alberto and {Arcodia}, Riccardo and {Ryu}, Taeho},
        title = "{Quasi-periodic eruptions from impacts between the secondary and a rigidly precessing accretion disc in an extreme mass-ratio inspiral system}",
      journal = {\aap},
         year = 2023,
        month = jul,
       volume = {675},
          eid = {A100},
        pages = {A100},
          doi = {10.1051/0004-6361/202346565},
archivePrefix = {arXiv},
       eprint = {2304.00775},
 primaryClass = {astro-ph.HE},
       adsurl = {https://ui.adsabs.harvard.edu/abs/2023A&A...675A.100F}
}

@ARTICLE{Giustini20,
       author = {{Giustini}, Margherita and {Miniutti}, Giovanni and {Saxton}, Richard D.},
        title = "{X-ray quasi-periodic eruptions from the galactic nucleus of RX J1301.9+2747}",
      journal = {\aap},
         year = 2020,
        month = apr,
       volume = {636},
          eid = {L2},
        pages = {L2},
          doi = {10.1051/0004-6361/202037610},
archivePrefix = {arXiv},
       eprint = {2002.08967},
 primaryClass = {astro-ph.HE},
       adsurl = {https://ui.adsabs.harvard.edu/abs/2020A&A...636L...2G}
}

@ARTICLE{Guolo25a,
       author = {{Guolo}, M. and {Mummery}, A. and {Wevers}, T. and {Nicholl}, M. and {Gezari}, S. and {Ingram}, A. and {Pasham}, D.~R.},
        title = "{The properties of GSN 069 accretion disk from a joint X-ray and UV spectral analysis: stress-testing quasi-periodic eruption models}",
      journal = {arXiv e-prints},
         year = 2025,
        month = jan,
          eid = {arXiv:2501.03333},
        pages = {arXiv:2501.03333},
          doi = {10.48550/arXiv.2501.03333},
archivePrefix = {arXiv},
       eprint = {2501.03333},
 primaryClass = {astro-ph.HE},
       adsurl = {https://ui.adsabs.harvard.edu/abs/2025arXiv250103333G}
}

@ARTICLE{Guolo25b,
       author = {{Guolo}, M. and {Mummery}, A. and {Ingram}, A. and {Nicholl}, M. and {Gezari}, S. and {Nathan}, E.},
        title = "{A Time-dependent Solution for GSN 069 Disk Evolution and the Nature of Long-lived Tidal Disruption Events}",
      journal = {\apj},
         year = 2025,
        month = oct,
       volume = {992},
       number = {1},
          eid = {114},
        pages = {114},
          doi = {10.3847/1538-4357/ae039e},
archivePrefix = {arXiv},
       eprint = {2504.20148},
 primaryClass = {astro-ph.HE},
       adsurl = {https://ui.adsabs.harvard.edu/abs/2025ApJ...992..114G}
}

@ARTICLE{Guolo25c,
       author = {{Guolo}, M. and {Mummery}, A. and {van Velzen}, S. and {Gezari}, S. and {Nicholl}, M. and {Yao}, Y. and {Karmen}, M. and {Ajay}, Y. and {Wevers}, T. and {LeBaron}, N. and {Chornock}, R.},
        title = "{Compact Accretion Disks in the Aftermath of Tidal Disruption Events: Parameter Inference from Joint X-ray Spectra and UV/Optical Photometry Fitting}",
      journal = {arXiv e-prints},
         year = 2025,
        month = oct,
          eid = {arXiv:2510.26774},
        pages = {arXiv:2510.26774},
          doi = {10.48550/arXiv.2510.26774},
archivePrefix = {arXiv},
       eprint = {2510.26774},
 primaryClass = {astro-ph.HE},
       adsurl = {https://ui.adsabs.harvard.edu/abs/2025arXiv251026774G}
}

@ARTICLE{Mummery25b,
       author = {{Mummery}, Andrew and {Nathan}, Edward and {Ingram}, Adam and {Gardner}, M.},
        title = "{Fitting transients with discs (FITTED): a public light curve and spectral fitting package based on evolving relativistic discs}",
      journal = {\mnras},
         year = 2025,
        month = dec,
       volume = {544},
       number = {2},
        pages = {2225-2240},
          doi = {10.1093/mnras/staf1565},
archivePrefix = {arXiv},
       eprint = {2408.15048},
 primaryClass = {astro-ph.HE},
       adsurl = {https://ui.adsabs.harvard.edu/abs/2025MNRAS.544.2225M}
}

@ARTICLE{Linial23b,
       author = {{Linial}, Itai and {Metzger}, Brian D.},
        title = "{EMRI + TDE = QPE: Periodic X-Ray Flares from Star-Disk Collisions in Galactic Nuclei}",
      journal = {\apj},
         year = 2023,
        month = nov,
       volume = {957},
       number = {1},
          eid = {34},
        pages = {34},
          doi = {10.3847/1538-4357/acf65b},
archivePrefix = {arXiv},
       eprint = {2303.16231},
 primaryClass = {astro-ph.HE},
       adsurl = {https://ui.adsabs.harvard.edu/abs/2023ApJ...957...34L}
}

@ARTICLE{Linial25,
       author = {{Linial}, Itai and {Metzger}, Brian D. and {Quataert}, Eliot},
        title = "{QPEs from EMRI Debris Streams Impacting Accretion Disks in Galactic Nuclei}",
      journal = {\apj},
         year = 2025,
        month = oct,
       volume = {991},
       number = {2},
          eid = {147},
        pages = {147},
          doi = {10.3847/1538-4357/adfa0e},
archivePrefix = {arXiv},
       eprint = {2506.10096},
 primaryClass = {astro-ph.HE},
       adsurl = {https://ui.adsabs.harvard.edu/abs/2025ApJ...991..147L}
}

@ARTICLE{Linial24a,
       author = {{Linial}, Itai and {Quataert}, Eliot},
        title = "{Period evolution of repeating transients in galactic nuclei}",
      journal = {\mnras},
         year = 2024,
        month = jan,
       volume = {527},
       number = {2},
        pages = {4317-4329},
          doi = {10.1093/mnras/stad3470},
archivePrefix = {arXiv},
       eprint = {2309.15849},
 primaryClass = {astro-ph.HE},
       adsurl = {https://ui.adsabs.harvard.edu/abs/2024MNRAS.527.4317L}
}

@ARTICLE{Linial24c,
       author = {{Linial}, Itai and {Metzger}, Brian D.},
        title = "{Coupled Disk-star Evolution in Galactic Nuclei and the Lifetimes of QPE Sources}",
      journal = {\apj},
         year = 2024,
        month = oct,
       volume = {973},
       number = {2},
          eid = {101},
        pages = {101},
          doi = {10.3847/1538-4357/ad639e},
archivePrefix = {arXiv},
       eprint = {2404.12421},
 primaryClass = {astro-ph.HE},
       adsurl = {https://ui.adsabs.harvard.edu/abs/2024ApJ...973..101L}
}

@INPROCEEDINGS{ciao,
       author = {{Fruscione}, Antonella and {McDowell}, Jonathan C. and {Allen}, Glenn E. and {Brickhouse}, Nancy S. and {Burke}, Douglas J. and {Davis}, John E. and {Durham}, Nick and {Elvis}, Martin and {Galle}, Elizabeth C. and {Harris}, Daniel E. and {Huenemoerder}, David P. and {Houck}, John C. and {Ishibashi}, Bish and {Karovska}, Margarita and {Nicastro}, Fabrizio and {Noble}, Michael S. and {Nowak}, Michael A. and {Primini}, Frank A. and {Siemiginowska}, Aneta and {Smith}, Randall K. and {Wise}, Michael},
        title = "{CIAO: Chandra's data analysis system}",
    booktitle = {Observatory Operations: Strategies, Processes, and Systems},
         year = 2006,
       editor = {{Silva}, David R. and {Doxsey}, Rodger E.},
       series = {Society of Photo-Optical Instrumentation Engineers (SPIE) Conference Series},
       volume = {6270},
        month = jun,
          eid = {62701V},
        pages = {62701V},
          doi = {10.1117/12.671760},
       adsurl = {https://ui.adsabs.harvard.edu/abs/2006SPIE.6270E..1VF}
}

@ARTICLE{Miniutti19,
       author = {{Miniutti}, G. and {Saxton}, R.~D. and {Giustini}, M. and {Alexander}, K.~D. and {Fender}, R.~P. and {Heywood}, I. and {Monageng}, I. and {Coriat}, M. and {Tzioumis}, A.~K. and {Read}, A.~M. and {Knigge}, C. and {Gandhi}, P. and {Pretorius}, M.~L. and {Ag{\'\i}s-Gonz{\'a}lez}, B.},
        title = "{Nine-hour X-ray quasi-periodic eruptions from a low-mass black hole galactic nucleus}",
      journal = {\nat},
         year = 2019,
        month = sep,
       volume = {573},
       number = {7774},
        pages = {381-384},
          doi = {10.1038/s41586-019-1556-x},
archivePrefix = {arXiv},
       eprint = {1909.04693},
 primaryClass = {astro-ph.HE},
       adsurl = {https://ui.adsabs.harvard.edu/abs/2019Natur.573..381M}
}

@ARTICLE{Yao25a,
       author = {{Yao}, Philippe Z. and {Quataert}, Eliot and {Jiang}, Yan-Fei and {Lu}, Wenbin and {White}, Christopher J.},
        title = "{Star‑Disk Collisions: Implications for Quasi-periodic Eruptions and Other Transients near Supermassive Black Holes}",
      journal = {\apj},
         year = 2025,
        month = jan,
       volume = {978},
       number = {1},
          eid = {91},
        pages = {91},
          doi = {10.3847/1538-4357/ad8911},
archivePrefix = {arXiv},
       eprint = {2407.14578},
 primaryClass = {astro-ph.HE},
       adsurl = {https://ui.adsabs.harvard.edu/abs/2025ApJ...978...91Y}
}

@ARTICLE{Wevers25,
       author = {{Wevers}, Thomas and {Guolo}, Muryel and {Lockwood}, Sean and {Mummery}, Andrew and {Pasham}, Dheeraj R. and {Arcodia}, Riccardo},
        title = "{Time-resolved Hubble Space Telescope UV observations of an X-ray quasi-periodic eruption source}",
      journal = {arXiv e-prints},
         year = 2025,
        month = jan,
          eid = {arXiv:2501.03335},
        pages = {arXiv:2501.03335},
          doi = {10.48550/arXiv.2501.03335},
archivePrefix = {arXiv},
       eprint = {2501.03335},
 primaryClass = {astro-ph.HE},
       adsurl = {https://ui.adsabs.harvard.edu/abs/2025arXiv250103335W}
}

@ARTICLE{Miniutti23a,
       author = {{Miniutti}, G. and {Giustini}, M. and {Arcodia}, R. and {Saxton}, R.~D. and {Read}, A.~M. and {Bianchi}, S. and {Alexander}, K.~D.},
        title = "{Repeating tidal disruptions in GSN 069: Long-term evolution and constraints on quasi-periodic eruptions' models}",
      journal = {\aap},
         year = 2023,
        month = feb,
       volume = {670},
          eid = {A93},
        pages = {A93},
          doi = {10.1051/0004-6361/202244512},
archivePrefix = {arXiv},
       eprint = {2207.07511},
 primaryClass = {astro-ph.HE},
       adsurl = {https://ui.adsabs.harvard.edu/abs/2023A&A...670A..93M}
}

@ARTICLE{Quintin23,
       author = {{Quintin}, E. and {Webb}, N.~A. and {Guillot}, S. and {Miniutti}, G. and {Kammoun}, E.~S. and {Giustini}, M. and {Arcodia}, R. and {Soucail}, G. and {Clerc}, N. and {Amato}, R. and {Markwardt}, C.~B.},
        title = "{Tormund's return: Hints of quasi-periodic eruption features from a recent optical tidal disruption event}",
      journal = {\aap},
         year = 2023,
        month = jul,
       volume = {675},
          eid = {A152},
        pages = {A152},
          doi = {10.1051/0004-6361/202346440},
archivePrefix = {arXiv},
       eprint = {2306.00438},
 primaryClass = {astro-ph.HE},
       adsurl = {https://ui.adsabs.harvard.edu/abs/2023A&A...675A.152Q}
}

@ARTICLE{Tagawa23,
       author = {{Tagawa}, Hiromichi and {Haiman}, Zolt{\'a}n},
        title = "{Flares from stars crossing active galactic nucleus discs on low-inclination orbits}",
      journal = {\mnras},
         year = 2023,
        month = nov,
       volume = {526},
       number = {1},
        pages = {69-79},
          doi = {10.1093/mnras/stad2616},
archivePrefix = {arXiv},
       eprint = {2304.03670},
 primaryClass = {astro-ph.HE},
       adsurl = {https://ui.adsabs.harvard.edu/abs/2023MNRAS.526...69T}
}

@ARTICLE{Xian21,
       author = {{Xian}, Jingtao and {Zhang}, Fupeng and {Dou}, Liming and {He}, Jiasheng and {Shu}, Xinwen},
        title = "{X-Ray Quasi-periodic Eruptions Driven by Star-Disk Collisions: Application to GSN069 and Probing the Spin of Massive Black Holes}",
      journal = {\apjl},
         year = 2021,
        month = nov,
       volume = {921},
       number = {2},
          eid = {L32},
        pages = {L32},
          doi = {10.3847/2041-8213/ac31aa},
archivePrefix = {arXiv},
       eprint = {2110.10855},
 primaryClass = {astro-ph.HE},
       adsurl = {https://ui.adsabs.harvard.edu/abs/2021ApJ...921L..32X}
}

@ARTICLE{Baldini26,
       author = {{Baldini}, Pietro and {Rau}, Arne and {Merloni}, Andrea and {Trakhtenbrot}, Benny and {Arcodia}, Riccardo and {Giustini}, Margherita and {Miniutti}, Giovanni and {Brennan}, Se{\'a}n J. and {Freyberg}, Michael and {S{\'a}nchez-S{\'a}ez}, Paula and {Grotova}, Iuliia and {Liu}, Zhu and {Lian}, Tianying and {Nandra}, Kirpal},
        title = "{Discovery of crested quasi-periodic eruptions following the most luminous SRG/eROSITA tidal disruption event}",
      journal = {arXiv e-prints},
         year = 2026,
        month = feb,
          eid = {arXiv:2602.03932},
        pages = {arXiv:2602.03932},
          doi = {10.48550/arXiv.2602.03932},
archivePrefix = {arXiv},
       eprint = {2602.03932},
 primaryClass = {astro-ph.HE},
       adsurl = {https://ui.adsabs.harvard.edu/abs/2026arXiv260203932B}
}

@ARTICLE{Mummery23,
       author = {{Mummery}, Andrew and {Wevers}, Thomas and {Saxton}, Richard and {Pasham}, Dheeraj},
        title = "{From X-rays to physical parameters: a comprehensive analysis of thermal tidal disruption event X-ray spectra}",
      journal = {\mnras},
         year = 2023,
        month = mar,
       volume = {519},
       number = {4},
        pages = {5828-5847},
          doi = {10.1093/mnras/stac3798},
archivePrefix = {arXiv},
       eprint = {2301.07419},
 primaryClass = {astro-ph.HE},
       adsurl = {https://ui.adsabs.harvard.edu/abs/2023MNRAS.519.5828M}
}

@ARTICLE{Mummery25a,
       author = {{Mummery}, Andrew},
        title = "{Collisions with tidal disruption event disks: implications for quasi-periodic X-ray eruptions}",
      journal = {arXiv e-prints},
         year = 2025,
        month = apr,
          eid = {arXiv:2504.21456},
        pages = {arXiv:2504.21456},
          doi = {10.48550/arXiv.2504.21456},
archivePrefix = {arXiv},
       eprint = {2504.21456},
 primaryClass = {astro-ph.HE},
       adsurl = {https://ui.adsabs.harvard.edu/abs/2025arXiv250421456M}
}

@ARTICLE{Zhou24a,
       author = {{Zhou}, Cong and {Huang}, Lei and {Guo}, Kangrou and {Li}, Ya-Ping and {Pan}, Zhen},
        title = "{Probing orbits of stellar mass objects deep in galactic nuclei with quasiperiodic eruptions}",
      journal = {\prd},
         year = 2024,
        month = may,
       volume = {109},
       number = {10},
          eid = {103031},
        pages = {103031},
          doi = {10.1103/PhysRevD.109.103031},
archivePrefix = {arXiv},
       eprint = {2401.11190},
 primaryClass = {astro-ph.HE},
       adsurl = {https://ui.adsabs.harvard.edu/abs/2024PhRvD.109j3031Z}
}

@ARTICLE{Miniutti25,
       author = {{Miniutti}, G. and {Franchini}, A. and {Bonetti}, M. and {Giustini}, M. and {Chakraborty}, J. and {Arcodia}, R. and {Saxton}, R. and {Quintin}, E. and {Kosec}, P. and {Linial}, I. and {Sesana}, A.},
        title = "{Eppur si muove: Evidence of disc precession or a sub-milliparsec SMBH binary in the QPE-emitting galaxy GSN 069}",
      journal = {\aap},
         year = 2025,
        month = jan,
       volume = {693},
          eid = {A179},
        pages = {A179},
          doi = {10.1051/0004-6361/202452400},
archivePrefix = {arXiv},
       eprint = {2411.13460},
 primaryClass = {astro-ph.HE},
       adsurl = {https://ui.adsabs.harvard.edu/abs/2025A&A...693A.179M}
}

@ARTICLE{Zhou24b,
       author = {{Zhou}, Cong and {Zhong}, Binyu and {Zeng}, Yuhe and {Huang}, Lei and {Pan}, Zhen},
        title = "{Probing orbits of stellar mass objects deep in galactic nuclei with quasiperiodic eruptions. II. Population analysis}",
      journal = {\prd},
         year = 2024,
        month = oct,
       volume = {110},
       number = {8},
          eid = {083019},
        pages = {083019},
          doi = {10.1103/PhysRevD.110.083019},
archivePrefix = {arXiv},
       eprint = {2405.06429},
 primaryClass = {astro-ph.HE},
       adsurl = {https://ui.adsabs.harvard.edu/abs/2024PhRvD.110h3019Z}
}

@ARTICLE{Zhou25,
       author = {{Zhou}, Cong and {Zeng}, Yuhe and {Pan}, Zhen},
        title = "{Secular Evolution of Quasiperiodic Eruptions}",
      journal = {\apj},
         year = 2025,
        month = jun,
       volume = {985},
       number = {2},
          eid = {242},
        pages = {242},
          doi = {10.3847/1538-4357/adcee2},
archivePrefix = {arXiv},
       eprint = {2411.18046},
 primaryClass = {astro-ph.HE},
       adsurl = {https://ui.adsabs.harvard.edu/abs/2025ApJ...985..242Z}
}

@ARTICLE{Chakraborty24,
       author = {{Chakraborty}, Joheen and {Arcodia}, Riccardo and {Kara}, Erin and {Miniutti}, Giovanni and {Giustini}, Margherita and {Tetarenko}, Alexandra J. and {Rhodes}, Lauren and {Franchini}, Alessia and {Bonetti}, Matteo and {Burdge}, Kevin B. and {Goodwin}, Adelle J. and {Maccarone}, Thomas J. and {Merloni}, Andrea and {Ponti}, Gabriele and {Remillard}, Ronald A. and {Saxton}, Richard D.},
        title = "{Testing EMRI Models for Quasi-periodic Eruptions with 3.5 yr of Monitoring eRO-QPE1}",
      journal = {\apj},
         year = 2024,
        month = apr,
       volume = {965},
       number = {1},
          eid = {12},
        pages = {12},
          doi = {10.3847/1538-4357/ad2941},
archivePrefix = {arXiv},
       eprint = {2402.08722},
 primaryClass = {astro-ph.HE},
       adsurl = {https://ui.adsabs.harvard.edu/abs/2024ApJ...965...12C}
}

@ARTICLE{Schmidt02,
       author = {{Schmidt}, W.},
        title = "{Celestial mechanics in Kerr spacetime}",
      journal = {Classical and Quantum Gravity},
         year = 2002,
        month = may,
       volume = {19},
       number = {10},
        pages = {2743-2764},
          doi = {10.1088/0264-9381/19/10/314},
archivePrefix = {arXiv},
       eprint = {gr-qc/0202090},
 primaryClass = {gr-qc},
       adsurl = {https://ui.adsabs.harvard.edu/abs/2002CQGra..19.2743S}
}

@ARTICLE{Foreman-Mackey13,
       author = {{Foreman-Mackey}, Daniel and {Hogg}, David W. and {Lang}, Dustin and {Goodman}, Jonathan},
        title = "{emcee: The MCMC Hammer}",
      journal = {\pasp},
         year = 2013,
        month = mar,
       volume = {125},
       number = {925},
        pages = {306},
          doi = {10.1086/670067},
archivePrefix = {arXiv},
       eprint = {1202.3665},
 primaryClass = {astro-ph.IM},
       adsurl = {https://ui.adsabs.harvard.edu/abs/2013PASP..125..306F}
}

@ARTICLE{Calzetti00,
       author = {{Calzetti}, Daniela and {Armus}, Lee and {Bohlin}, Ralph C. and {Kinney}, Anne L. and {Koornneef}, Jan and {Storchi-Bergmann}, Thaisa},
        title = "{The Dust Content and Opacity of Actively Star-forming Galaxies}",
      journal = {\apj},
         year = 2000,
        month = apr,
       volume = {533},
       number = {2},
        pages = {682-695},
          doi = {10.1086/308692},
archivePrefix = {arXiv},
       eprint = {astro-ph/9911459},
 primaryClass = {astro-ph},
       adsurl = {https://ui.adsabs.harvard.edu/abs/2000ApJ...533..682C}
}

@ARTICLE{Done12,
       author = {{Done}, Chris and {Davis}, S.~W. and {Jin}, C. and {Blaes}, O. and {Ward}, M.},
        title = "{Intrinsic disc emission and the soft X-ray excess in active galactic nuclei}",
      journal = {\mnras},
         year = 2012,
        month = mar,
       volume = {420},
       number = {3},
        pages = {1848-1860},
          doi = {10.1111/j.1365-2966.2011.19779.x},
archivePrefix = {arXiv},
       eprint = {1107.5429},
 primaryClass = {astro-ph.HE},
       adsurl = {https://ui.adsabs.harvard.edu/abs/2012MNRAS.420.1848D}
}

@ARTICLE{Kaaz23,
       author = {{Kaaz}, Nicholas and {Liska}, Matthew T.~P. and {Jacquemin-Ide}, Jonatan and {Andalman}, Zachary L. and {Musoke}, Gibwa and {Tchekhovskoy}, Alexander and {Porth}, Oliver},
        title = "{Nozzle Shocks, Disk Tearing, and Streamers Drive Rapid Accretion in 3D GRMHD Simulations of Warped Thin Disks}",
      journal = {\apj},
         year = 2023,
        month = sep,
       volume = {955},
       number = {1},
          eid = {72},
        pages = {72},
          doi = {10.3847/1538-4357/ace051},
archivePrefix = {arXiv},
       eprint = {2210.10053},
 primaryClass = {astro-ph.HE},
       adsurl = {https://ui.adsabs.harvard.edu/abs/2023ApJ...955...72K}
}

@ARTICLE{Huang25,
       author = {{Huang}, Xiaoshan and {Linial}, Itai and {Jiang}, Yan-Fei},
        title = "{Multiband Emission from Star─Disk Collisions and Implications for Quasiperiodic Eruptions}",
      journal = {\apj},
         year = 2025,
        month = nov,
       volume = {993},
       number = {2},
          eid = {186},
        pages = {186},
          doi = {10.3847/1538-4357/ae07ca},
archivePrefix = {arXiv},
       eprint = {2506.11231},
 primaryClass = {astro-ph.HE},
       adsurl = {https://ui.adsabs.harvard.edu/abs/2025ApJ...993..186H}
}

@ARTICLE{Yao26,
       author = {{Yao}, Philippe Z. and {Quataert}, Eliot and {Jiang}, Yan-Fei and {Linial}, Itai},
        title = "{Star-Disk Collisions II: Debris Stream Dynamics and Implications for QPEs and Other Transients Near SMBHs}",
      journal = {arXiv e-prints},
         year = 2026,
        month = jul,
          eid = {arXiv:2607.08823},
        pages = {arXiv:2607.08823},
          doi = {10.48550/arXiv.2607.08823},
archivePrefix = {arXiv},
       eprint = {2607.08823},
 primaryClass = {astro-ph.HE},
       adsurl = {https://ui.adsabs.harvard.edu/abs/2026arXiv260708823Y}
}
\bibliographystyle{aasjournal}

\appendix
\counterwithin{figure}{section}

\section{Derivation of the timing equation}\label{app:timing_derivation}
A disk collision occurs where the secondary crosses the local midplane of the
accretion disk.  For the EMRI orbit part of the problem we follow textbook Keplerian notation, and assume that over one orbit the full GR expression can be well approximated as a Newtonian ellipse. This approximation scheme is a statement regarding the ratio $r_g/a$, the inverse of the semi-major axis in gravitational radii, which we discuss further below.  This means we have an  
oriented ellipse specified by the inclination \(i\), ascending-node
longitude \(\Omega\), and argument of pericenter \(\omega\), with position
along the orbit given by the argument of latitude \(u\) (the in-plane angle
from the ascending node) and true anomaly \(f=u-\omega\).  We reproduce the
standard Keplerian orbital expressions here only to fix the sign convention.  In a frame with
\(\hat{\mathbf z}\) along the black hole's spin axis the unit position vector of the EMRI is
\begin{equation}
\label{eq:rhat}
\hat{\mathbf r}(u)=
\begin{pmatrix}
\cos\Omega\cos u-\sin\Omega\sin u\cos i\\
\sin\Omega\cos u+\cos\Omega\sin u\cos i\\
\sin u\,\sin i
\end{pmatrix}.
\end{equation}
The local disk midplane at the crossing radius is a flat ring through the
origin whose unit normal is tilted from the spin axis by the local warp angle
\(\beta\) and twisted to the Bardeen--Petterson azimuth \(\gamma\),
\begin{equation}
\label{eq:nhat}
\hat{\mathbf n}=\bigl(\sin\beta\cos\gamma,\ \sin\beta\sin\gamma,\ \cos\beta\bigr).
\end{equation}
The signed height above the midplane vanishes (i.e., the EMRI crosses the disk) when
\(\hat{\mathbf r}\cdot\hat{\mathbf n}=0\).  Expanding the dot product and
collecting coefficients of \(\cos u\) and \(\sin u\), the two azimuths enter
only through the single difference
\(\psi\equiv\gamma-\Omega\):
\begin{equation}
\label{eq:cross}
\hat{\mathbf r}\cdot\hat{\mathbf n}
=\sin\beta\,\cos\psi\,\cos u
+\bigl(\cos\beta\,\sin i+\sin\beta\,\cos i\,\sin\psi\bigr)\sin u
=0.
\end{equation}
This follows from the angle-difference identities
\(\cos\gamma\cos\Omega+\sin\gamma\sin\Omega=\cos\psi\) and
\(\sin\gamma\cos\Omega-\cos\gamma\sin\Omega=\sin\psi\).
Equation~(\ref{eq:cross}) is of the form
\(\mathcal{A}\cos u+\mathcal{B}\sin u=0\), or equivalently
\({\cal R}\sin(u+\chi)=0\) with \({\cal R}=\sqrt{\mathcal{A}^2+\mathcal{B}^2}\) and
\(\tan\chi=\mathcal{A}/\mathcal{B}\).  The crossings therefore sit at
\(u=-\chi\pmod{\pi}\), where \(\chi\) is the extra orbital phase beyond the
node required to reach the disk.  It vanishes identically for a flat,
spin-aligned disk (\(\beta=0\)), which pins the crossings to the orbital
nodes \(u=0,\pi\).  
The two roots half a turn apart are the two crossings per orbit.  Tracking
one node, \(u_n=\pi-\chi_n\), the corresponding true anomaly (measured from
pericenter) is \(f_n=u_n-\omega_n=\pi-\omega_n-\chi_n\).

We stress again that we do not use full general relativity for the orbital problem.  The approximation we employ 
is a scale separation between the orbital period and the apsidal
precession period (with the nodal period guaranteed to be longer still), which is a statement of how relativistic the orbit is:
\(P/P_{\rm aps}\sim r_g/a\ll 1\).  Over a single orbit the orientation may
then be held frozen and Kepler's laws used in full, after which the
relativistic advance of pericenter and line of nodes is applied as a discrete ``kick'' to the orbital elements. We show in \cite{Mummery26c} that this approximation is good to sub-percent levels for the types of orbits considered here.

The time from pericenter to a given true anomaly is likewise the classical
Kepler time of flight (Kepler's equation)
\begin{equation}
\label{eq:tof}
\tau(f)=\frac{P}{2\pi}\bigl(E-e\sin E\bigr),\qquad
\tan\frac{E}{2}=\sqrt{\frac{1-e}{1+e}}\,\tan\frac{f}{2},
\end{equation}
with \(E\) the eccentric anomaly and \(P\) the mean orbital period.  The
\(n\)th crossing of the chosen node therefore arrives at
\(t_n=t_{\rm peri}+nP+\tau(f_n)\).  Isolating the equation of the center
\(E-e\sin E-f\) and substituting \(f_n\) gives
\begin{equation}
\label{eq:tnsplit}
t_n=t_{\rm peri}+nP+\frac{P}{2\pi}\bigl(E_n-e\sin E_n-f_n\bigr)
+\frac{P}{2\pi}\bigl(\pi-\omega_n-\chi_n\bigr).
\end{equation}
An overall epoch and every term linear in the orbit number \(n\)---including
the secular advance of \(\omega_n\)---are absorbed into a reference ephemeris
\(t_0+nP\).  What remains of the geometric offset is the nodal contribution
\(-(P/2\pi)\chi_n\), which cannot be absorbed: it is set by the instantaneous
misalignment of the orbital and disk planes.  The crossing times are therefore
\begin{align}
t_n=t_0+nP+\frac{P}{2\pi}&\bigl(E_n-e\sin E_n-f_n\bigr)-\frac{P}{2\pi}\,\chi_n,
\label{eq:timing_app}
\\
\tan\frac{E_n}{2}&=\sqrt{\frac{1-e}{1+e}}\,\tan\frac{f_n}{2}, \\
f_n & =\pi-\omega_n-\chi_n,
\end{align}
where \(E_n\) is the classical eccentric anomaly, and the final term results
from the different planes of the orbit and disk:
\begin{equation}
    \chi_n = \arctan\!\left[\frac{\sin\beta\,\cos\psi_n}
    {\cos\beta\,\sin i + \sin\beta\,\cos i\,\sin\psi_n}\right].
    \label{eq:chi_app}
\end{equation}
Here \(\omega_n\) and \(\psi_n\) are the argument of pericenter and the
relative nodal phase at the \(n\)th crossing.  The second node is the same
sequence with \(\chi_n\to\chi_n+\pi\).

\section{Supplementary figures} \label{appendix}

\begin{figure*}
    \includegraphics[width=\textwidth]{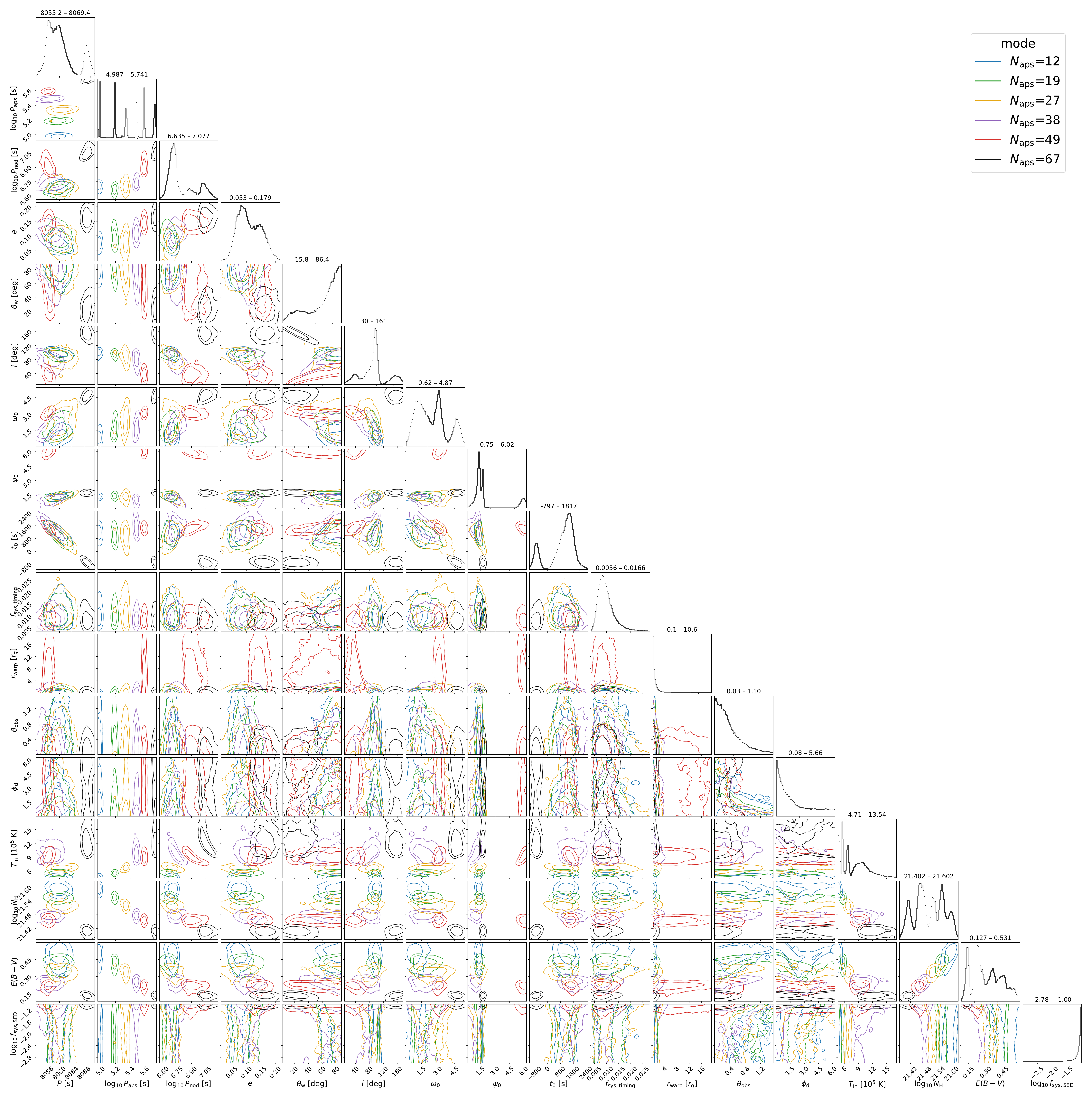}
    \caption{Full corner plot of the eRO-QPE2 joint spectral-timing fit in the sampled (period-basis) parametrization, combined over the six timing modes. Shaded contours enclose 68\% and 95\% of the posterior mass. The numbers above each marginal give the 90\% range.}
    \label{fig:ero_corner_full}
\end{figure*}

\begin{figure*}
    \includegraphics[width=\textwidth]{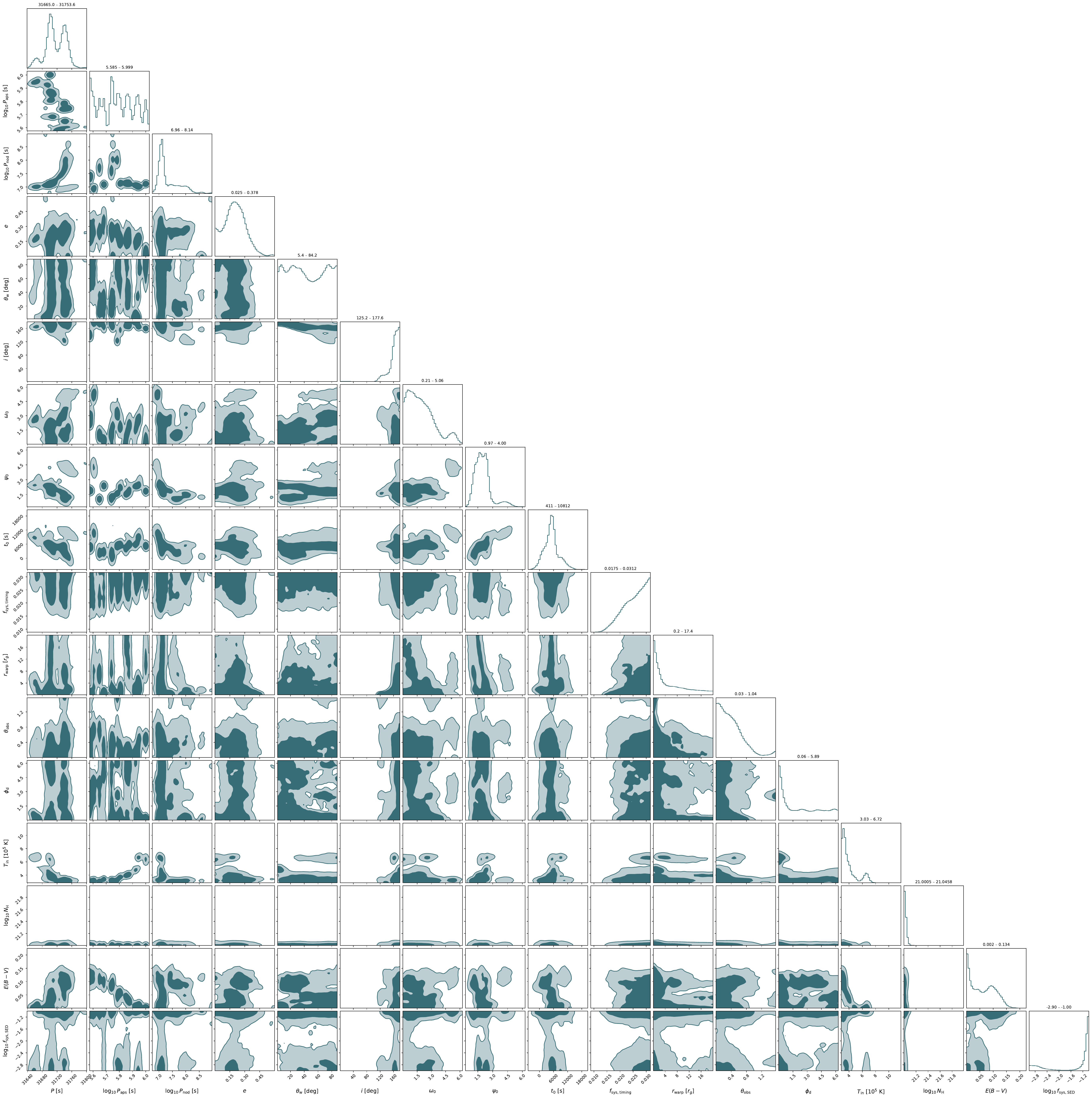}
    \caption{Full corner plot of the GSN~069 joint spectral-timing fit in the sampled (period-basis) parametrization. Shaded contours enclose 68\% and 95\% of the posterior mass. The numbers above each marginal give the 90\% range.}
    \label{fig:gsn_corner_full}
\end{figure*}

In Fig.~\ref{fig:ero_corner_full} we show the full corner plot of sampled parameters in eRO-QPE2, in the timing basis rather than the physical one (see Section~\ref{subsec:timing} for the distinction). In Fig.~\ref{fig:gsn_corner_full} we show the same for GSN~069.

\end{document}